\documentclass[a4paper,11pt]{article}
\usepackage[ruled,vlined,linesnumbered,noresetcount]{algorithm2e}
\usepackage[a4paper, total={6in, 8in}]{geometry}
\usepackage{color}
\usepackage{amssymb}
\usepackage{amsmath}
\usepackage{bbm}

\usepackage[nodisplayskipstretch]{setspace}

\usepackage[colorinlistoftodos,textwidth=1.7cm, textsize=footnotesize]{todonotes}

\usepackage{lineno}

\newcommand*\patchAmsMathEnvironmentForLineno[1]{%
	\expandafter\let\csname old#1\expandafter\endcsname\csname #1\endcsname
	\expandafter\let\csname oldend#1\expandafter\endcsname\csname end#1\endcsname
	\renewenvironment{#1}%
	{\linenomath\csname old#1\endcsname}%
	{\csname oldend#1\endcsname\endlinenomath}}%
\newcommand*\patchBothAmsMathEnvironmentsForLineno[1]{%
	\patchAmsMathEnvironmentForLineno{#1}%
	\patchAmsMathEnvironmentForLineno{#1*}}%
\AtBeginDocument{%
	\patchBothAmsMathEnvironmentsForLineno{equation}%
	\patchBothAmsMathEnvironmentsForLineno{align}%
	\patchBothAmsMathEnvironmentsForLineno{flalign}%
	\patchBothAmsMathEnvironmentsForLineno{alignat}%
	\patchBothAmsMathEnvironmentsForLineno{gather}%
	\patchBothAmsMathEnvironmentsForLineno{multline}%
}

\usepackage[english]{babel}
\usepackage{amsthm}
\newtheorem{remark}{Remark}
\usepackage{float}
\usepackage{natbib}
\usepackage{bibentry}
\usepackage{multirow}
\usepackage{graphicx}
\usepackage{comment}
\usepackage{authblk}
\usepackage{caption} 
\newtheorem{assumption}{Assumption}
\newtheorem{theorem}{Theorem}

\newtheorem{lemma}{Lemma}

\newtheorem{define}{Definition}

\providecommand{\keywords}[1]{\textbf{\textit{Keyword:}} #1}

\begin{document}
\title{Monte Carlo Co-Ordinate Ascent Variational Inference}
\author[1]{Lifeng Ye}
\author[1]{Alexandros Beskos}
\author[1,2]{Maria De Iorio}
\author[3]{Jie Hao}

\affil[1]{Department of Statistical Science, University College London}
\affil[2]{Yale-NUS College, Singapore}
\affil[3]{Key Laboratory of Systems Biomedicine (Ministry of Education), Shanghai Center for Systems Biomedicine, Shanghai Jiao Tong University}

\affil[ ]{\textit {lifeng.ye.13@ucl.ac.uk, a.beskos@ucl.ac.uk, maria@yale-nus.edu.sg, j.hao@sjtu.edu.cn}}

\date{}                     %
\setcounter{Maxaffil}{0}
\renewcommand\Affilfont{\itshape\small}
\maketitle

\begin{abstract}
In Variational Inference (VI), coordinate-ascent  and gradient-based approaches are two major types of algorithms for approximating difficult-to-compute probability densities. In real-world implementations of complex models, Monte Carlo methods are widely used to estimate expectations in coordinate-ascent approaches and gradients in derivative-driven ones. We discuss a Monte Carlo Co-ordinate Ascent VI (MC-CAVI) algorithm that makes use of Markov chain Monte Carlo (MCMC) methods in the calculation of  expectations required within Co-ordinate Ascent VI (CAVI). We show that, under regularity conditions, an MC-CAVI recursion will get arbitrarily close to a maximiser of the evidence lower bound (ELBO) with any  given high probability. In numerical examples, the performance of MC-CAVI algorithm is compared with that of MCMC and -- as a representative of derivative-based VI methods -- of Black Box VI (BBVI). We discuss
and demonstrate MC-CAVI's suitability for models with \emph{hard constraints} in simulated and real examples. We compare MC-CAVI's performance with that of MCMC in an important complex model used in Nuclear Magnetic Resonance (NMR) spectroscopy data analysis -- BBVI is nearly impossible to be employed in this setting due to the hard constraints involved in the model.
\end{abstract}


\section{Introduction}
Variational Inference (VI) \citep{Jordan1999, wainwright_jordan_2014} is a powerful method to approximate intractable integrals. As an alternative strategy to Markov chain Monte Carlo (MCMC) sampling, VI is fast, relatively straightforward for monitoring convergence and typically easier to scale to large data \citep{bleivi} than MCMC.  The key idea of VI is to approximate difficult-to-compute conditional densities of latent variables, given observations, via use of optimization. A family of distributions is assumed for the latent variables, as an approximation to the exact conditional distribution. VI aims at finding the member, amongst the selected family, that minimizes the Kullback-Leibler (KL) divergence from the conditional law of interest. 

Let $x$ and $z$ denote, respectively, the observed data and latent variables. The goal of the inference problem is to identify the conditional density (assuming a relevant reference measure, e.g.~Lebesgue) of latent variables given observations, i.e. $p(z| x)$.
Let $\mathcal{L}$ denote a family of densities defined over the space of latent variables -- we denote members of this family as $q=q(z)$ below. The goal of VI is to find the element of the family closest in KL divergence to the true $p(z| x)$. Thus, the original inference problem can be rewritten as an optimization one: identify $q^*$ such that 
\begin{equation}
\label{eq:min}
q^* = \operatornamewithlimits{argmin}\limits_{q\in \mathcal{L}}\textrm{KL}(q\mid p(\cdot| x))
\end{equation}
for the KL-divergence defined as
\begin{align*}
\textrm{KL}(q\mid p(\cdot | x)) &= \mathbb{E}_{q}[\log q(z)] - \mathbb{E}_q[\log p(z| x)] \\ &= \mathbb{E}_q[\log q(z)] - \mathbb{E}_q[\log p(z,x)] + \log p(x),
\end{align*}
with $\log p(x)$ being constant w.r.t.~$z$. Notation $\mathbb{E}_q$ refers to   expectation taken over $z\sim q$. Thus, minimizing the KL divergence is equivalent to maximising the evidence lower bound, ELBO$(q)$, given by
\begin{equation}
\textrm{ELBO}(q) = \mathbb{E}_q[\log p(z,x)] - \mathbb{E}_q[\log q(z)].
\label{elbo1}
\end{equation}
Let $\mathsf{S}_p\subseteq \mathbb{R}^{m}$, $m\ge 1$, denote the support of the target $p(z|x)$, and 
$\mathsf{S}_{q}\subseteq \mathbb{R}^{m}$ the support of a variational density $q\in\mathcal{L}$ -- assumed to be common over all members
$q\in\mathcal{L}$. Necessarily, $\mathsf{S}_p\subseteq \mathsf{S}_q$, otherwise the KL-divergence will diverge to $+\infty$.

Many VI algorithms focus on the mean-field variational family, where variational densities in $\mathcal{L}$ are assumed to factorise over blocks of $z$. That is, 
\begin{equation}
\label{eq:meanfield}
q(z) = \prod_{i=1}^b q_i(z_i),\quad \mathsf{S}_q = \mathsf{S}_{q_{1}}\times \cdots \times \mathsf{S}_{q_{b}}, \quad
z=(z_1,\ldots, z_{b})\in \mathsf{S}_q, \,\,\,z_i\in \mathsf{S}_{q_{i}},
\end{equation}
for individual supports $\mathsf{S}_{q_{i}}\subseteq\mathbb{R}^{m_i}$, $m_i\ge 1$, $1\le i\le b$, for some $b\ge 1$, and $\sum_{i}m_i =m$.
It is advisable that highly correlated latent variables are placed in the same block to improve the 
performance of the VI method.

There are, in general, two types of approaches to maximise ELBO in VI: a co-ordinate ascent approach and a gradient-based one. Co-ordinate ascent VI (CAVI) \citep{bishop_2006} is amongst the most commonly used algorithms in this context. To obtain a local maximiser for ELBO, CAVI sequentially optimizes each factor of the mean-field variational density, while holding the others fixed. Analytical 
calculations on function space -- involving variational derivatives -- imply that, 
for given fixed $q_1,\ldots, q_{i-1},q_{i+1},\ldots, q_b$, 
ELBO$(q)$ is maximised for 
\begin{equation}
\label{eq:recursion}
q_i(z_i)\propto \exp\big\{\mathbb{E}_{-i}[\log p(z_{i_-},z_{i},z_{i_+},x)]\big\},
\end{equation}
\noindent
where $z_{-i}:=(z_{i_-},z_{i_+})$ denotes vector $z$ having removed component $z_i$, 
with ${i_-}$ (resp.~${i_+}$) denoting the ordered indices that are smaller (resp.~larger) than~$i$; $\mathbb{E}_{-i}$ is the expectation taken under $z_{-i}$ following its  variational distribution, denoted $q_{-i}$.
The above suggest immediately an iterative algorithm, 
guaranteed to provide values for ELBO$(q)$ that cannot decrease as the updates are carried out. 

The expected value
$\mathbb{E}_{-i}[\log p(z_{i_-},z_{i},z_{i_+},x)]$ can be difficult to derive analytically. 
Also, CAVI typically requires traversing the entire dataset at each iteration, which can be overly computationally expensive for large datasets.
Gradient-based approaches, which can potentially scale up to large data -- alluding here to recent Stochastic-Gradient-type methods -- can be an effective alternative for ELBO optimisation. However, such algorithms
have their own challenges, e.g.
in the case reparameterization Variational Bayes (VB) analytical derivation of gradients of the log-likelihood can often be problematic, while in the case of score-function VB the requirement of the gradient of $\log q$ restricts the range of the family $\mathcal{L}$ we can choose from.

In real-world applications, hybrid methods combining Monte Carlo with recursive algorithms are common, e.g., Auto-Encoding Variational Bayes, 
Doubly-Stochastic Variational Bayes for non-conjugate inference, Stochastic Expectation-Maximation (EM) \citep{Beaumont2025, Sisson1760, mcem}. In VI, Monte Carlo is often used to estimate the expectation within CAVI or the gradient within derivative-driven methods.
This is the case, e.g.,  for Stochastic VI \citep{svi} and Black-Box VI (BBVI) \citep{bbvi}.

BBVI is used in this work as a representative of gradient-based VI algorithms. It allows carrying out VI over a wide range of complex models. The variational density $q$ is typically chosen within a parametric family, so finding $q^*$ 
in~(\ref{eq:min}) is equivalent to determining an optimal set of parameters that characterize $q_i=q_i(\cdot|\lambda_i)$, $\lambda_{i}\in \Lambda_i\subseteq \mathbb{R}^{d_i}$, $1\le d_i$, $1\le i\le b$, with $\sum_{i=1}^{b}d_i=d$. The gradient of ELBO w.r.t.~the variational parameters $\lambda=(\lambda_1,\ldots,\lambda_b)$ equals  
\begin{equation}
\label{eq:mainBB}
\nabla_{\lambda} \textrm{ELBO}(q) := \mathbb{E}_q\big[\nabla_{\lambda}\log q(z| \lambda)\{\log p(z,x)-\log q(z| \lambda)\}\big]
\end{equation}
and can be approximated by black-box Monte Carlo estimators as, e.g., 
\begin{equation}
\label{eq:BBVIest}
\widehat{\nabla_{\lambda}\textrm{ELBO}(q)} := \tfrac{1}{N}\sum^N_{n=1}\big[\nabla_{\lambda}\log q(z^{(n)}| \lambda)\{\log p(z^{(n)},x)-\log q(z^{(n)}|\lambda)\}\big],
\end{equation}
with $z^{(n)} \stackrel{iid}{\sim} q(z| \lambda)$, $1\le n\le N$,  $N\ge 1$. The approximated gradient of ELBO can then be used within a stochastic optimization procedure to update $\lambda$ at the $k$th iteration with
\begin{equation}
\lambda_{k+1} \leftarrow \lambda_k + \rho_k \widehat{\nabla_{\lambda_k}\textrm{ELBO}(q)},
\label{eq:BBVIit}
\end{equation}
where $\{\rho_k\}_{k\ge 0}$ is a Robbins-Monro-type step-size sequence \citep{robbins1951}.
As we will see in later sections, BBVI is accompanied by generic 
variance reduction methods, as the variability of (\ref{eq:BBVIest}) 
for complex models can be large.

\begin{remark}[Hard Constraints]
\label{rem:issue}
Though gradient-based VI methods are some times more straightforward to apply than co-ordinate ascent ones, -- e.g.~combined with the use of modern approaches for automatic differentiation \citep{advi} -- co-ordinate ascent methods can still be important for models with \emph{hard constraints}, where gradient-based algorithms are laborious to apply. (We adopt the viewpoint here that one chooses variational densities that respect the constaints of the target, for improved accuracy.) Indeed, notice in the brief description we have given above for CAVI and BBVI, the two methodologies are structurally different, as CAVI does not necessarily require to be build up via the introduction of an exogenous variational parameter $\lambda$. Thus, in the context of a support for the target $p(z|x)$ that involves complex constraints, 
a CAVI approach overcomes this issue naturally by blocking together the $z_i$'s responsible for the constraints. In contrast, introduction of the variational parameter $\lambda$ creates sometimes severe 
complications in the development of the derivative-driven algorithm, as normalising constants that depend on $\lambda$ are extremely difficult to calculate analytically
and obtain their derivatives. Thus, a main argument spanning this work -- and illustrated within it -- 
is that co-ordinate-ascent-based VI methods have a critical role to play amongst VI approaches for important classes of statistical models. 
\end{remark}

The main contributions of the paper are:
\begin{itemize}
\item[(i)]
We discuss, and then apply a Monte Carlo CAVI (MC-CAVI) algorithm in a sequence of problems of increasing complexity, and study its performance. As the name suggests, MC-CAVI 
 uses the Monte Carlo principle for the approximation of the difficult-to-compute conditional expectations, $\mathbb{E}_{-i}[\log p(z_{i_-},z_{i},z_{i_+},x)]$, within CAVI. 
\item[(ii)]
We provide a justification for the algorithm by showing analytically that, under suitable regularity conditions, MC-CAVI will get arbitrarily close to a maximiser of the ELBO with high probability. 

\item[(iii)] We contrast MC-CAVI with MCMC and BBVI through simulated and real examples, some of which involve hard constraints; we demonstrate MC-CAVI's effectiveness in an important application imposing such hard constraints, with real data 
in the context of Nuclear Magnetic Resonance (NMR) spectroscopy.
\end{itemize}

\begin{remark}
Inserting Monte Carlo steps within a VI approach (that might use a mean field or another approximation) is not uncommon in the VI literature. E.g., 
\cite{forb:07} employ an MCMC procedure in the context of a Variational EM (VEM), to obtain estimates of the normalizing constant for Markov Random Fields -- they provide asymptotic results for the correctness of the complete algorithm;
\cite{tran:16} apply Mean-Field Variational Bayes (VB) 
for Generalised Linear Mixed Models, and use Monte Carlo
for the approximation of analytically intractable required expectations under the variational densities; 
several references for related works are given in the above papers.
Our work focuses on MC-CAVI, and develops theory that is appropriate for this VI method. This algorithm has \emph{not} been studied analytically in the literature, thus the development of its theoretical justification  -- even if it borrows elements from Monte Carlo EM -- is new.
\end{remark}

The rest of the paper is organised as follows. 
Section \ref{sec:MCCAVI} presents briefly the MC-CAVI algorithm. It also provides -- in a specified setting -- an analytical result illustrating non-accumulation of Monte Carlo errors in the execution of the recursions of the algorithm. That is, with a probability arbitrarily close to 1, the variational solution provided by MC-CAVI can be as close as required to the one of CAVI, for a big enough Monte Carlo sample size, regardless of the number of algorithmic iterations.
Section \ref{sec:numerics} shows two numerical examples, contrasting MC-CAVI with alternative algorithms. 
Section \ref{sec:nmr} presents an implementation of MC-CAVI in a real, complex, challenging posterior distribution arising in metabolomics. This is a practical application, involving hard constraints, chosen to illustrate the potential of MC-CAVI in this context. We finish with some conclusions in Section \ref{sec:discussion}.

\section{MC-CAVI Algorithm}
\label{sec:MCCAVI}

\subsection{Description of the Algorithm}
\label{subsec:CAVI}

We begin with a description of the basic CAVI algorithm.
A double subscript will be used to identify block variational densities: $q_{i,k}(z_i)$ (resp.~$q_{-i,k}(z_{-i})$) will refer to the density of the $i$th block (resp.~all blocks but the $i$th), after $k$ updates have been carried out on that block density (resp.~$k$ updates have been carried out on the blocks preceding the $i$th, and $k-1$ updates on the blocks following the $i$th).
\begin{itemize}
\item Step 0:  Initialize probability density functions $q_{i,0}(z_i)$, $i=1,\ldots, b$.
\item Step $k$: 
 For $k\ge 1$, given $q_{i,k-1}(z_i)$, $i=1,\ldots, b$, 
 execute: 
 \begin{itemize}
 \item For $i=1,\ldots, b$, update:
\begin{align*}
 \log q_{i,k}(z_i) = const. + \mathbb{E}_{-i,k}[\log p(z,x)],
\end{align*}
with $\mathbb{E}_{-i,k}$ taken 
w.r.t.~$z_{-i}\sim q_{-i,k}$. 
 \end{itemize}
\item Iterate until convergence.
\end{itemize}

\noindent 
Assume that the expectations $\mathbb{E}_{-i}[\log p(z,x)]$, $\{i:i\in\mathcal{I}\}$, for an index set $\mathcal{I}\subseteq\{1,\ldots, b\}$, 
can be obtained analytically, over all updates of the variational density $q(z)$; and that this is not the case for $i\notin\mathcal{I}$. Intractable integrals can be approximated via a Monte Carlo method. (As we will see in the applications in the sequel, such a Monte Carlo device typically uses samples from an appropriate MCMC algorithm.)
In particular, for $i\notin \mathcal{I}$, one obtains $N\ge 1$ samples from the current $q_{-i}(z_{-i})$  and uses the standard Monte Carlo estimate 
\begin{equation*} 
 \widehat{\mathbb{E}}_{-i}[\log p(z_{i_-},z_{i},z_{i_+},x)]
 = \frac{\sum_{n=1}^{N} \log p(z_{i_-}^{(n)},z_{i},z_{i_+}^{(n)},x)}{N}.
\end{equation*}
 Implementation of such an approach gives rise to MC-CAVI, 
 described in Algorithm~\ref{MC-CAVI}.
\begin{algorithm}[!h]
\SetAlgoLined
\vspace{0.2cm}
\SetKwInOut{Input}{Require}
\SetKw{KwBy}{by}
\Input{Number of iterations $T$.\vspace{0.2cm}}
\Input{Number of Monte Carlo samples $N$.\vspace{0.2cm}}
\Input{$\mathbb{E}_{-i} [\log  p(z_{i_-},z_i, z_{i_+},x)]$ in closed form, for $i\in \mathcal{I}$.\vspace{0.2cm}}
  Initialize $q_{i,0}(z_i)$, $i=1,\ldots, b$.\vspace{0.2cm} \\
 \For{$k= 1:T$\vspace{0.2cm} }{
    \For{$i=1:b$\vspace{0.2cm} }{
    If $i\in\mathcal{I}$, set
     $q_{i,k}(z_i) \propto \exp \big\{ \mathbb{E}_{-i,k}[\log p(z{_{i_-}},z_i, z{_{i_+}},x)] \big\} $ \vspace{0.2cm} \;  
    If $i\notin\mathcal{I}$:\\
    Obtain $N$ samples, $(z_{i_{-},k}^{(n)},z_{i_{+},k-1}^{(n)})$, $1\le n \le N$, from 
    $q_{-i,k}(z_{-i})$.
    \\
    Set $$q_{i,k}(z_i) \propto  \exp \big\{ \tfrac{\sum_{n=1}^{N} \textrm{log}\; p(z_{i_-,k}^{(n)},z_{i},z_{i_+,k-1}^{(n)},x)}{N} \big\}.$$
    }}
\caption{MC-CAVI}\label{MC-CAVI}
\end{algorithm}
\hfill \\

\subsection{Applicability of MC-CAVI}

We discuss here the class of problems for which MC-CAVI can be applied. 
It is desirable to avoid settings where the order of samples or statistics to be stored 
in memory increases with the iterations of the algorithm. 
To set-up the ideas we begin with CAVI itself.  Motivated by the 
standard exponential class of distributions, we work as follows. 

Consider the case when the target density $p(z,x)\equiv f(z)$ -- we omit reference to the data $x$ in what follows, as $x$ is fixed and irrelevant for our purposes (notice that $f$ is not required to integrate to $1$) --  is assumed to have the structure, 
\begin{align}
\label{eq:class}
f(z) = h(z)\exp\big\{ \langle \eta, T(z) \rangle - A(\eta)       \big\},\quad z\in \mathsf{S}_p,
\end{align}
for $s$-dimensional constant vector $\eta=(\eta_1,\ldots, \eta_s)$, vector function $T(z)=(T_1(z),\ldots, T_{s}(z))$, with some $s\ge 1$, and relevant scalar functions $h>0$, $A$; $\langle \cdot,\cdot \rangle$ is the standard inner product in $\mathbb{R}^{s}$. 
Also, we are given the choice of block-variational densities $q_1(z_1),\ldots, q_b(z_b)$ in (\ref{eq:meanfield}). Following the definition of CAVI from Section \ref{subsec:CAVI} -- 
assuming that the algorithm can be applied, i.e.~all required expectations can be obtained analytically --
the number of `sufficient' statistics, say $T_{i,k}$ giving rise to the definition of $q_{i,k}$ 
will always be upper bounded by $s$. Thus, in our working scenario, CAVI will be applicable with 
a computational cost that is upper bounded by a constant within the class of target distributions in
(\ref{eq:class}) -- assuming relevant costs for calculating expectations remain bounded over the algorithmic iterations.   

Moving on to MC-CAVI, following the definition of index set $\mathcal{I}$ in Section \ref{subsec:CAVI},
recall that a Monte Carlo approach is required when updating $q_i(z_i)$ for $i\notin \mathcal{I}$, $1\le i \le b$. In such a scenario, controlling computational costs amounts to having a target (\ref{eq:class}) admitting the factorisations, 
\begin{equation}
\label{eq:fact}
h(z) \equiv h_i(z_i)h_{-i}(z_{-i}),\quad T_{l}(z) \equiv T_{l,i}(z_{i})T_{l,-i}(z_{-i}), \,\,\,1\le l\le s,
\quad\,\, \textrm{for all }\,i\notin \mathcal{I}.
\end{equation}
Once (\ref{eq:fact}) is satisfied, we do not need to store all $N$ samples from $q_{-i}(z_{-i})$, but simply some relevant averages keeping the cost per iteration for the algorithm bounded. We stress that the combination of characterisations in (\ref{eq:class})-(\ref{eq:fact}) is very general and will typically be satisfied for most practical statistical models.

\subsection{Theoretical Justification of  MC-CAVI}

An advantageous feature of MC-CAVI versus derivative-driven VI methods is its structural similarity with Monte Carlo Expectation-Maximization (MCEM). Thus, one can build on results in the MCEM literature to prove asymptotical properties of MC-CAVI; see e.g.~\cite{mc-em, boot:99, levi:01, fort:03}.
To avoid technicalities related with working on general spaces of probability density functions, we begin by assuming a parameterised setting for the variational densities -- as in the BBVI case -- 
with the family of variational densities being closed under CAVI or (more generally) MC-CAVI updates.

\begin{assumption}[Closedness of Parameterised $q(\cdot)$ Under Variational Update]
\label{ass:family}

For the CAVI or the MC-CAVI algorithm, each $q_{i,k}(z_i)$ density obtained during the iterations of the algorithm, $1\leq i\leq b$, $k\ge 0$, is of the parametric form
$$q_{i,k}(z_i) = q_i(z_i|\lambda_{i}^{k}),$$ for a unique $\lambda_{i}^{k}\in \Lambda_i\subseteq \mathbb{R}^{d_i}$, for some  $d_i\ge 1$, for all $1\le i \le b$. \\
(Let $d=\sum_{i=1}^b {d_i}$ and $\Lambda =\Lambda_1 \times \cdots \times \Lambda_b $.)

\end{assumption}
\noindent Under Assumption \ref{ass:family}, CAVI and MC-CAVI can be corresponded to some well-defined maps  
$M:\Lambda\mapsto\Lambda$, $\mathcal{M}_N:\Lambda\mapsto\Lambda$ respectively, so that, 
given current variational parameter $\lambda$, one step of the algorithms can be expressed in terms of 
a new parameter $\lambda'$ (different for each case) obtained via the updates
\begin{equation*}
\textrm{CAVI:}\,\,\,\,\lambda' = M(\lambda); \qquad  \textrm{MC-CAVI:}\,\,\,\,\lambda' = 
\mathcal{M}_N(\lambda).
\end{equation*}
\indent  For an analytical study of the convergence properties of CAVI itself and relevant regularity conditions, see e.g.~\cite[Proposition 2.7.1]{bert:99},
or numerous other resources in numerical optimisation. 
Expressing the MC-CAVI update -- say, the $(k+1)$th one -- as
\begin{equation}
\label{eq:perturb}
\lambda^{k+1} = M(\lambda^k) + \{ \mathcal{M}_N(\lambda^k) - M(\lambda^k) \}, 
\end{equation}
it can be seen as a random perturbation of a CAVI step. In the rest of this section we will explore the asymptotic properties of MC-CAVI. We follow closely the approach in \cite{mc-em} --  as it provides a less technical procedure, compared e.g.~to \cite{fort:03} or other works about MCEM -- making all appropriate adjustments to fit the derivations into the setting of the MC-CAVI methodology along the way. We denote by $M^{k}$, $\mathcal{M}_N^{k}$, the $k$-fold composition of $M$, $\mathcal{M}_{N}$ respectively, for $k\ge 0$.
\begin{assumption}	
\label{ass:regular}		
$\Lambda$ is an open subset of $\mathbb{R}^{d}$, and the  
mappings $\lambda\mapsto \textrm{ELBO}(q(\lambda))$, $\lambda\mapsto M(\lambda)$ are continuous on $\Lambda$.
\end{assumption}
\noindent If $M(\lambda)=\lambda$ for some $\lambda\in \Lambda$, then $\lambda$ is a fixed point of $M()$.
A given $\lambda^*\in \Lambda$ is called an isolated local maximiser of the ELBO$(q(\cdot))$ if there is a neighborhood of 
$\lambda^*$ over which $\lambda^*$ is the unique maximiser of the ELBO$(q(\cdot))$.
\begin{assumption}[Properties of $M(\cdot)$ Near a Local Maximum]	
\label{ass:M}
Let $\lambda^*\in\Lambda$ be an isolated local maximum of ELBO$(q(\cdot))$. Then,  
\begin{itemize}
\item[(i)] $\lambda^*$ is a fixed point of $M(\cdot)$;
\item[(ii)]there is a neighborhood $V\subseteq \Lambda$ of $\lambda^*$ over which $\lambda^*$ is a unique maximum, such that
 $\textrm{ELBO}(q(M(\lambda)))>\textrm{ELBO}(q(\lambda))$ for any $\lambda\in V\backslash\{\lambda^*\}$.
\end{itemize}
\end{assumption}
\noindent 
Notice that the above assumption refers to the deterministic update $M(\cdot)$, which performs co-ordinate ascent; thus requirements (i), (ii) are fairly weak for such a recursion.
The critical technical assumption required for delivering the convergence results in the rest of this section is the following one.
\begin{assumption}[Uniform Convergence in Probability on Compact Sets]
\label{ass:technical}
For any compact set $C\subseteq\Lambda$ the following holds: for any $\varrho,\varrho'>0$, there exists a positive integer $N_0$,
such that for all $N\ge N_0$ we have,
\begin{equation*}
\inf_{\lambda\in C} \mathrm{Prob}\,\big[\,  \big|  \mathcal{M}_N(\lambda)-M(\lambda)     \big| < \varrho \,  \big] 
> 1-\varrho' .
\end{equation*}
\end{assumption}

\noindent  
It is beyond the context of this paper to examine Assumption \ref{ass:technical} in more depth. We will only stress that Assumption \ref{ass:technical} is the sufficient structural condition
that allows to extend closeness between CAVI and MC-CAVI updates in a single algorithmic step into 
one for arbitrary number of steps.
   
We continue with a definition.
\begin{define}
\label{def:stable}
A fixed point $\lambda^*$ of $M(\cdot)$ is said to be asymptotically stable if,
\begin{itemize}
\item[(i)] for any neighborhood $V_1$ of $\lambda^*$, there is a neighborhood $V_2$ of $\lambda^*$ such that for all~$k\ge 0$ and all $\lambda\in V_2$, $M^k(\lambda)\in V_1$; 
\item[(ii)] there exists a neighbourhood $V$ of $\lambda^*$ such that
$\lim_{k\rightarrow\infty}M^k(\lambda)=\lambda^*$ if $\lambda\in V$.
\end{itemize}
\end{define}

We will state the main asymptotic result for MC-CAVI in Theorem \ref{th:stable} that follows; first we require Lemma
\ref{lem:stable}.

\begin{lemma}
\label{lem:stable}
Let Assumptions \ref{ass:family}-\ref{ass:M} hold.
If $\lambda^*$ is an isolated local maximiser of $\textrm{ELBO}(q(\cdot))$, then $\lambda^*$ is an asymptotically stable fixed point of $M(\cdot)$.
\end{lemma}

The main result of this section is as follows.

\begin{theorem} 
\label{th:stable}
Let Assumptions \ref{ass:family}-\ref{ass:technical} hold and $\lambda^*$ be an isolated local maximiser of $\mathrm{ELBO}(q(\cdot))$. Then there exists a neighbourhood, say $V_1$, of $\lambda^*$ such that for starting values
$\lambda\in V_1$ of  MC-CAVI algorithm and for all $\epsilon_1>0$, there exists a $k_0$ such that 
\begin{equation*}
\lim_{N\rightarrow \infty}\mathrm{Prob}\,\big(\,|\mathcal{M}_N^{k}-\lambda^* | < \epsilon_1 \textrm{ for some } k\leq k_0\,\big)= 1.
\end{equation*}
\end{theorem}

\noindent The proofs of Lemma \ref{lem:stable} and Theorem \ref{th:stable} can be found in Appendices \ref{sec:lem} and \ref{sec:theorem}, respectively.

\subsection{Stopping Criterion and Sample Size}

The method requires the specification of the Monte Carlo size $N$ and a stopping rule.

\subsubsection*{Principled - but Impractical - Approach}

As the algorithm approaches a local maximum, changes in ELBO should be getting closer to zero.
To evaluate the performance of MC-CAVI, one could, in principle, attempt to monitor the evolution of  ELBO during the algorithmic iterations.
For current variational distribution $q=(q_1,\ldots, q_b)$, assume that MC-CAVI is about to update $q_i$
with $q'_i= q'_{i,N}$, where the addition of the second subscript at this point emphasizes the dependence of the new value for $q_i$ on the Monte Carlo size $N$. Define,
\begin{equation*}
 \Delta\mathrm{ELBO}(q, N) = \mathrm{ELBO}(q_{i-},q'_{i,N},q_{i+}) - \mathrm{ELBO}(q). 
\end{equation*}
If the algorithm is close to a local maximum,  $\Delta$ELBO$(q, N)$ should be close to zero, at least for sufficiently large $N$. Given such a choice of $N$, an MC-CAVI recursion should be terminated once $\Delta$ELBO$(q, N)$ is smaller than a user-specified tolerance threshold.
 Assume that the random variable 
$\Delta$ELBO$(q, N)$ has mean $\mu = \mu(q, N)$ and variance $\sigma^2 = \sigma^2(q, N)$. 
Chebychev's inequality implies that, with probability  greater than or equal to $(1-1/K^2)$, $\Delta$ELBO$(q, N)$ lies within the interval $(\mu-K\sigma, \mu + K\sigma)$, for any real $K>0$. Assume that one fixes a large enough $K$.
The choice of $N$ and of a stopping criterion should be based on the requirements: 
\begin{itemize}
\item[(i)] $\sigma\leq \nu$, with $\nu$ a predetermined level of tolerance; 
\item[(ii)] the effective range $(\mu-K\sigma, \mu + K\sigma)$ should include zero, implying that $\Delta$ELBO$(q, N)$ differs from zero by less than $2K\sigma$. 
\end{itemize}
Requirement (i) provides a rule for the choice of $N$ -- assuming applied over all $1\le i \le  b$, for $q$ in areas close to a maximiser --, and requirement (ii) a rule for defining a stopping criterion. Unfortunately, the above considerations -- based on the proper term ELBO$(q)$ that VI aims to maximise --
involve quantities that are typically impossible to obtain analytically or via some reasonably expensive approximation.

\subsubsection*{Practical Considerations}
Similarly to MCEM, it is recommended that $N$ gets increased as the algorithm becomes more stable.
It is computationally inefficient to start with a large value of $N$ when the current variational distribution  can be far from the maximiser. In practice, one may monitor the convergence of the algorithm by plotting relevant \emph{statistics}  of the variational distribution versus the number of iterations. We can declare that convergence has been reached when such traceplots show relatively small random fluctuations (due to the Monte Carlo variability) around a fixed value.  At this point, one may terminate the algorithm or continue with a larger value of $N$, which will further decrease the traceplot variability. In the applications we encounter in the sequel, we typically have $N\le 100$, 
so calculating, for instance, Effective Sample Sizes to monitor the mixing performance of the MCMC steps is not practical.

\section{Numerical Examples -- Simulation Study}
\label{sec:numerics}

In this section we illustrate MC-CAVI with two simulated examples. 
First, we apply MC-CAVI and CAVI on a simple model to highlight main features and implementation strategies. 
Then,  we contrast MC-CAVI, MCMC, BBVI in a complex scenario with hard constraints.

\subsection{Simulated Example 1}
\label{sec:example1}
\begin{figure}
\begin{center}
\includegraphics[scale=0.35]{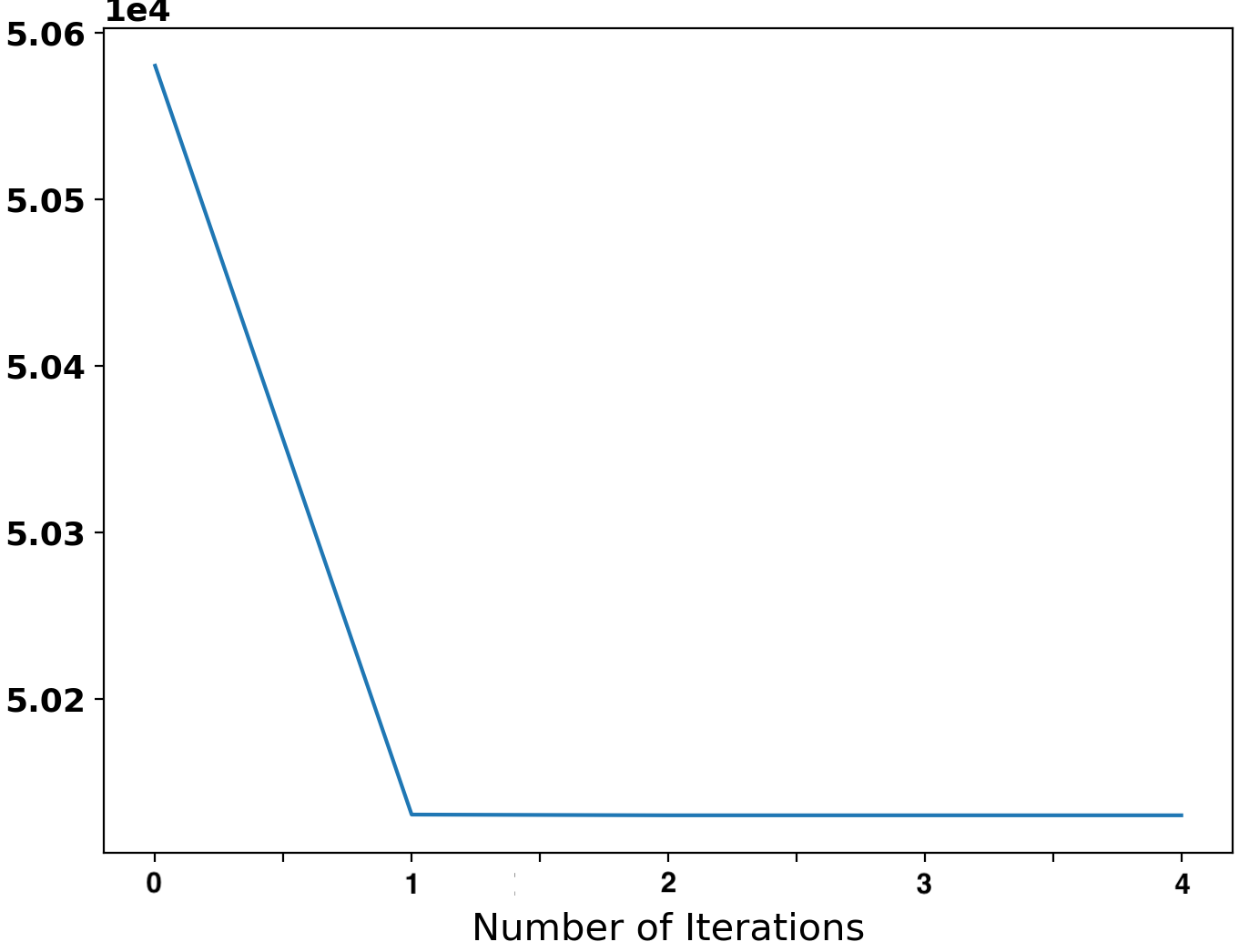}
\includegraphics[scale=0.35]{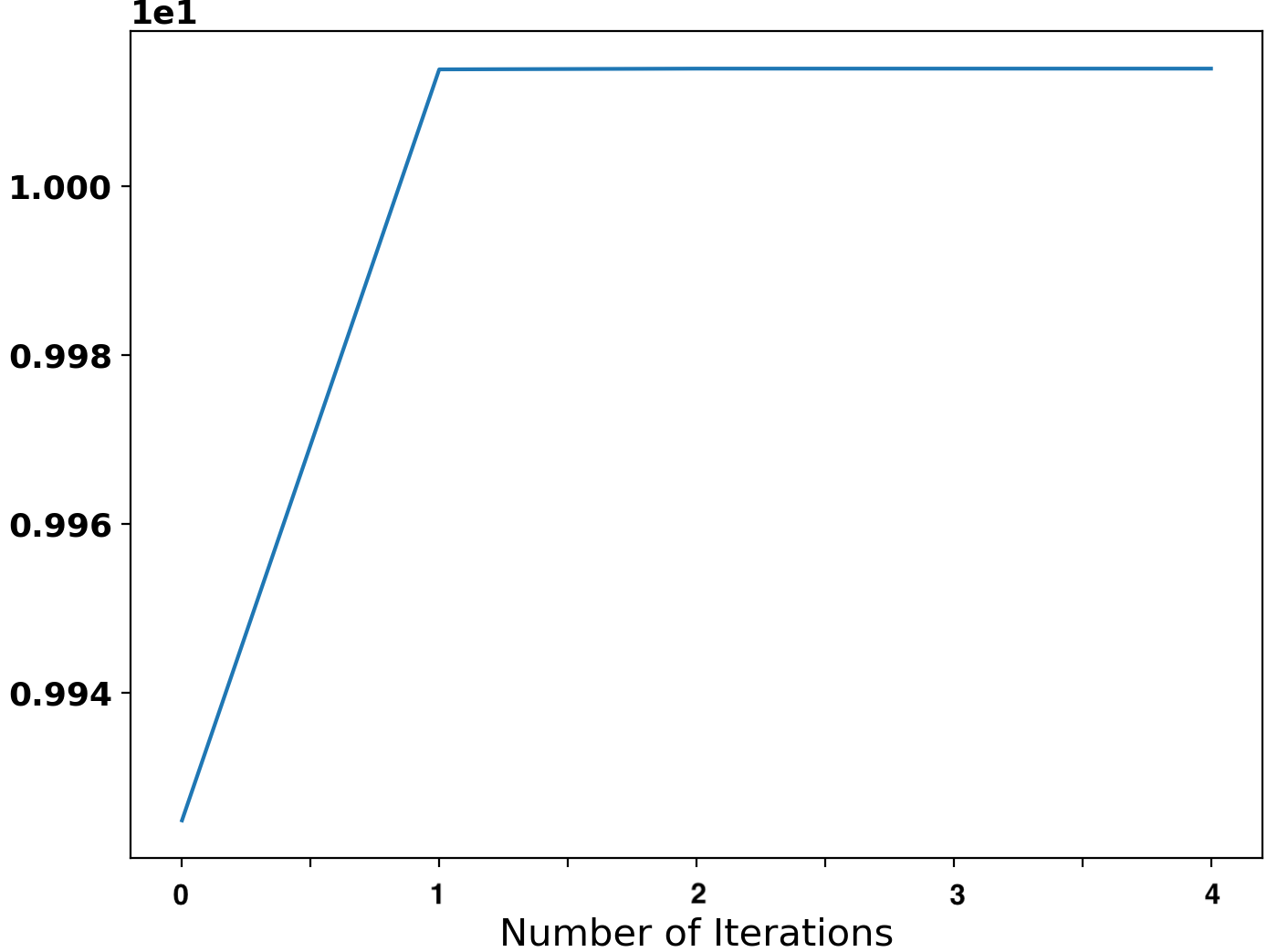}
\end{center}
\caption{Tracplots of $\zeta$ (left), $\theta$ (right) from application of CAVI on Simulated Example~1.}
\label{viresult1}
\end{figure}

We generate $n=10^3$ data points from $\mathrm{N}(10,100)$ and fit the semi-conjugate Bayesian model
\begin{align*}
\textrm{\underline{Example Model 1}} \\
{x_1, \ldots, x_n} &\sim \mathrm{N}(\vartheta,\tau^{-1}), \\
 \vartheta &\sim \mathrm{N}(0,\tau^{-1}), \\
 \tau &\sim \textrm{Gamma}(1,1).
 \end{align*}
Let $\bar{x}$ be the data sample mean. In each iteration, the CAVI density function -- see  (\ref{eq:recursion}) -- for $\tau$ is that of the Gamma distribution $\textrm{Gamma}(\tfrac{n+3}{2},\zeta)$, with
\begin{align*}
\zeta = 1 + \tfrac{(1+n)\mathbb{E}(\vartheta^2)-2(n\bar{x})\mathbb{E}(\vartheta)+\sum^n_{j=1}x^2_j}{2},
\end{align*} 
whereas for $\vartheta$ that of the normal distribution $\mathrm{N}(\frac{n\bar{x}}{1+n},\frac{1}{(1+n)\mathbb{E}(\tau)})$.
$(\mathbb{E}(\vartheta),\mathbb{E}(\vartheta^2))$ and $\mathbb{E}(\tau)$ denote the relevant expectations under the current CAVI distributions for $\vartheta$ and $\tau$ respectively; the former are initialized at 0 -- there is no need to initialise $\mathbb{E}(\tau)$ in this case. Convergence of CAVI can be monitored, e.g., via   
the sequence of values of $\theta := (1+n)\mathbb{E}(\tau)$ and $\zeta$. If the change in values of these two parameters is smaller than, say, $0.01\%$, we declare convergence. Figure \ref{viresult1} shows the traceplots of $\theta$, $\zeta$.
Convergence is reached within 0.0017secs\footnote{A Dell Latitude E5470 with Intel(R) Core(TM) i5-6300U CPU@2.40GHz is used for all experiments in this paper.}, after precisely two iterations, due to the simplicity of the model. The resulted CAVI distribution for $\vartheta$ is $\mathrm{N}(9.6,0.1)$, and for $\tau$ it is Gamma$(501.5,50130.3)$ so that  $\mathbb{E}(\tau) \approx 0.01$.

\begin{figure}
\begin{center}
\includegraphics[scale=0.5]{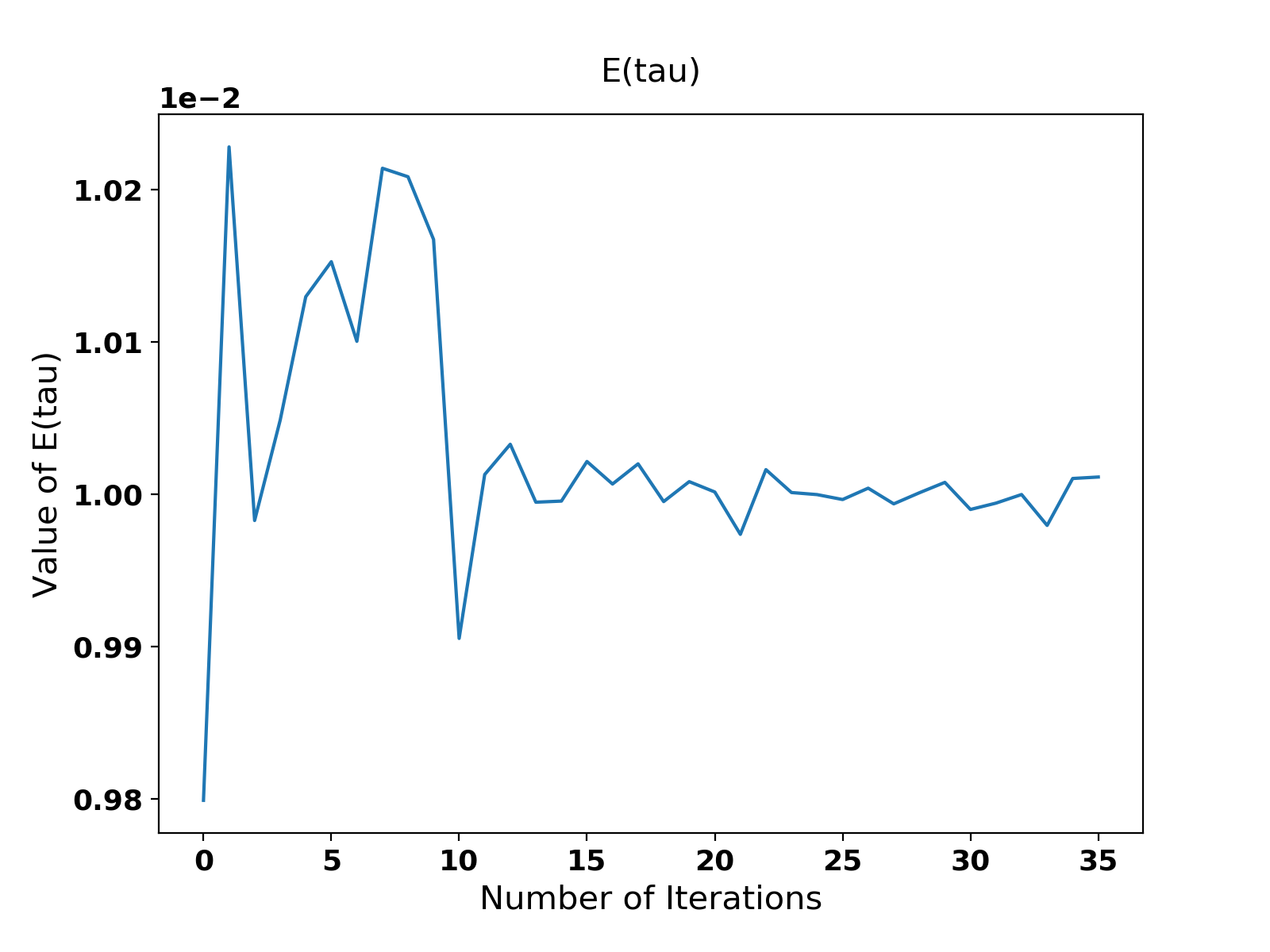}
\end{center}
\vspace{-0.4cm}
\caption{Traceplot of $\widehat{\mathbb{E}}(\tau)$ generated by MC-CAVI for Simulated Example 1, using $N=10$ for the first 10 iterations of the algorithm, and $N=10^3$ for the rest.}
\label{mcmcresult1}
\end{figure}

Assume now that $q(\tau)$ was intractable. 
Since $\mathbb{E}(\tau)$ is required to update the approximate distribution of $\vartheta$, an MCMC step can be employed to sample $\tau_1,\ldots, \tau_{N}$ from $q(\tau)$ to produce the Monte Carlo estimator $\widehat{\mathbb{E}}(\tau)=\sum^{N}_{j=1}\tau_j/N$. Within this MC-CAVI setting, $\widehat{\mathbb{E}}(\tau)$ will replace the exact 
${\mathbb{E}}(\tau)$ during the algorithmic iterations.
$(\mathbb{E}(\vartheta),\mathbb{E}(\vartheta^2))$ are initialised as in CAVI. 
For the first 10 iterations we set $N=10$, and for the remaining ones,  $N=10^3$ to reduce variability.
We monitor the values of  $\widehat{\mathbb{E}}(\tau)$ shown 
in Figure \ref{mcmcresult1}.  
The figure shows that MC-CAVI has stabilized after about $15$ iterations; algorithmic time was 0.0114secs. To remove some Monte Carlo variability, the final estimator of $\mathbb{E}(\tau)$ 
is produced by averaging the last 10 values of its traceplot, 
which gives $\widehat{\mathbb{E}}(\tau) = 0.01$, i.e.~a value very close to the one obtained by CAVI. The estimated distribution of $\vartheta$ is $\mathrm{N}(9.6,0.1)$, the same as with CAVI. 

\begin{figure}
\includegraphics[scale=0.35]{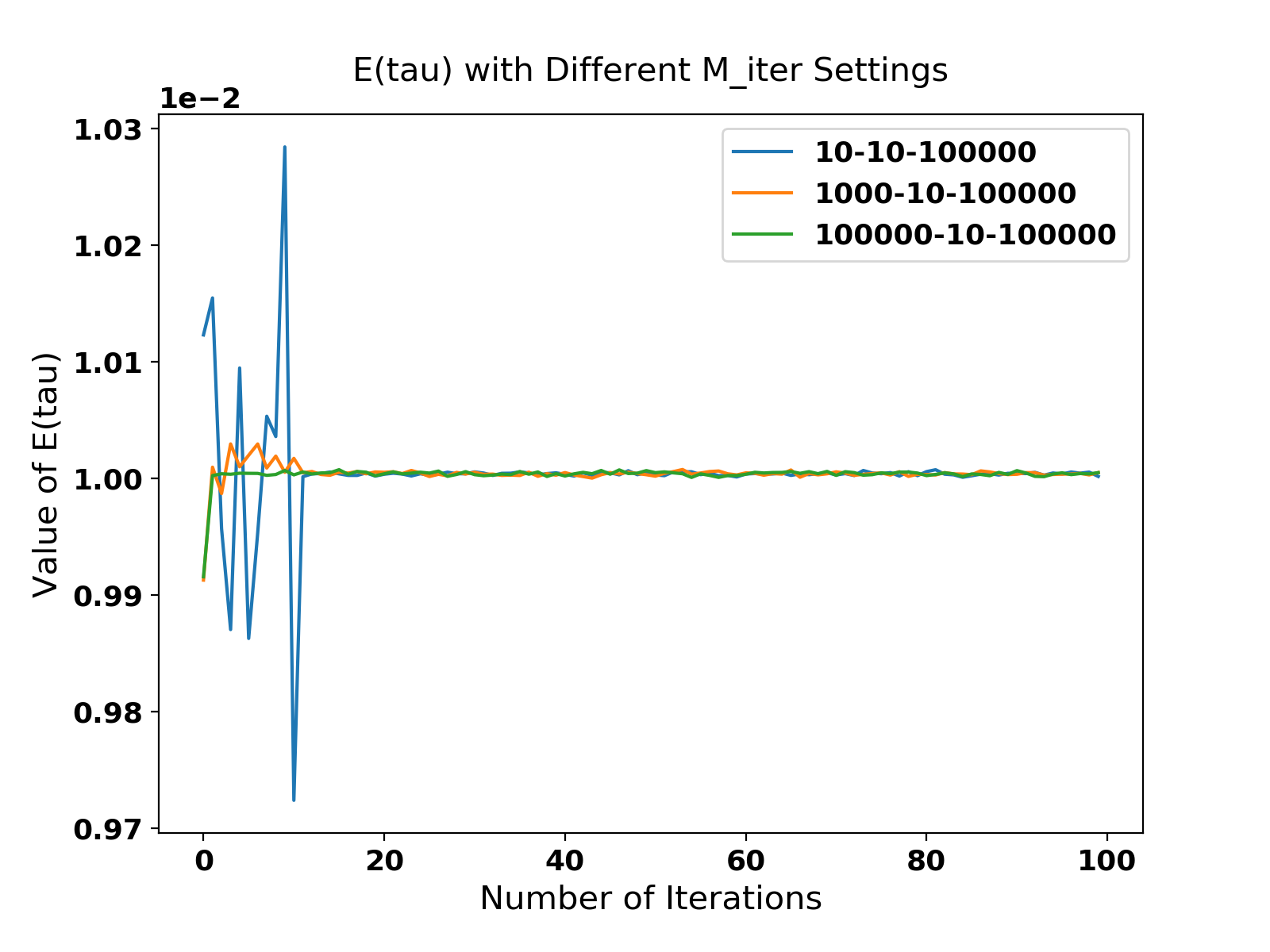}
\includegraphics[scale=0.35]{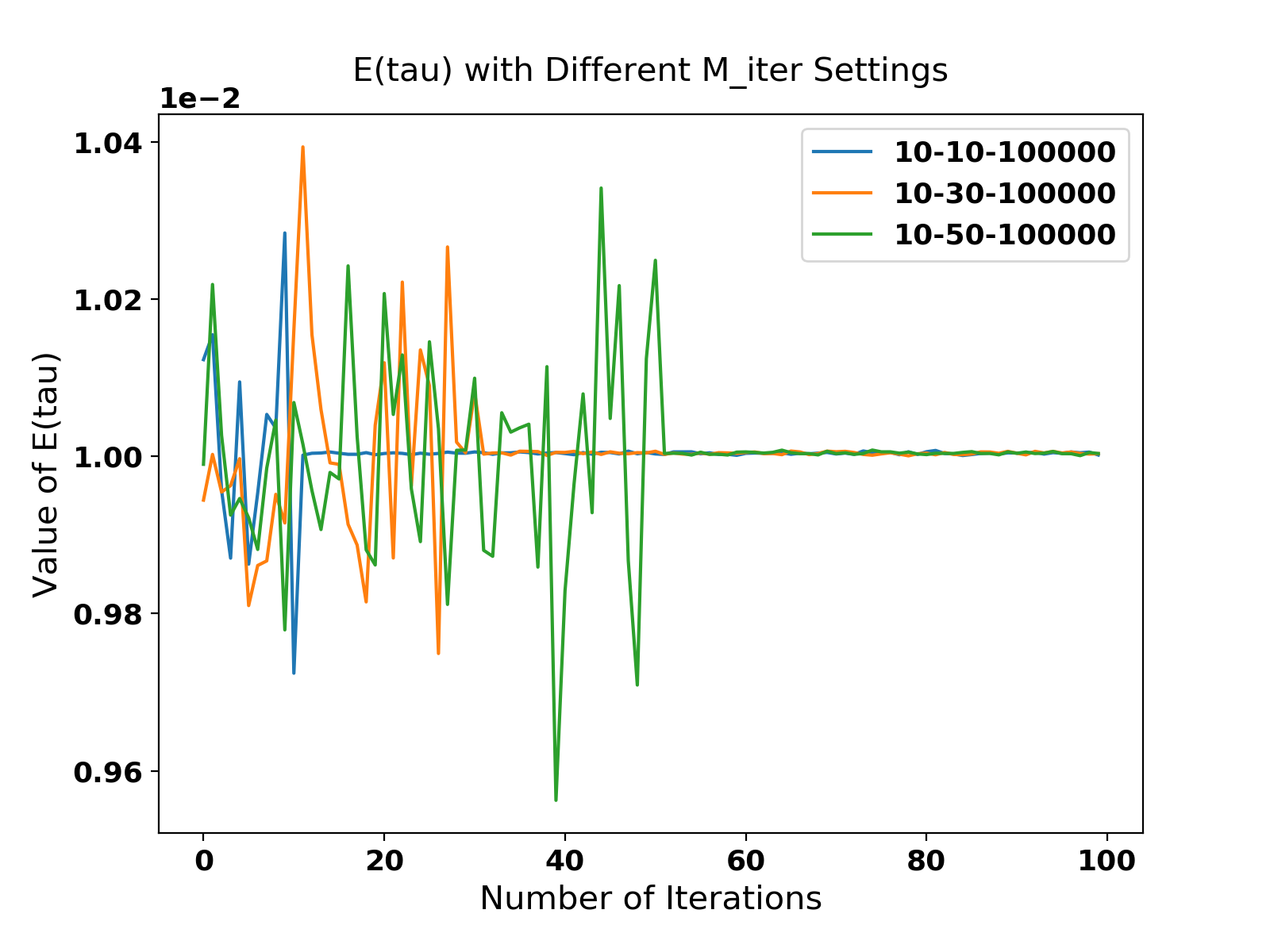}
\caption{Traceplot of $\widehat{\mathbb{E}}(\tau)$ under different settings of A-B-C (respectively, the value of $N$ in the burn-in period, the number of burn-in iterations and the value of $N$ after burn-in) for Simulated Example 1.}
\label{mitertune}
\end{figure}

The performance of MC-CAVI depends critically on the choice $N$. Let A be the value of $N$ in the burn-in period, B the number of burn-in iterations and  C the value of $N$ after burn-in. Figure \ref{mitertune} shows trace plots of $\widehat{\mathbb{E}}(\tau)$ under different settings of the triplet A-B-C.

\begin{figure}
\includegraphics[scale=0.45]{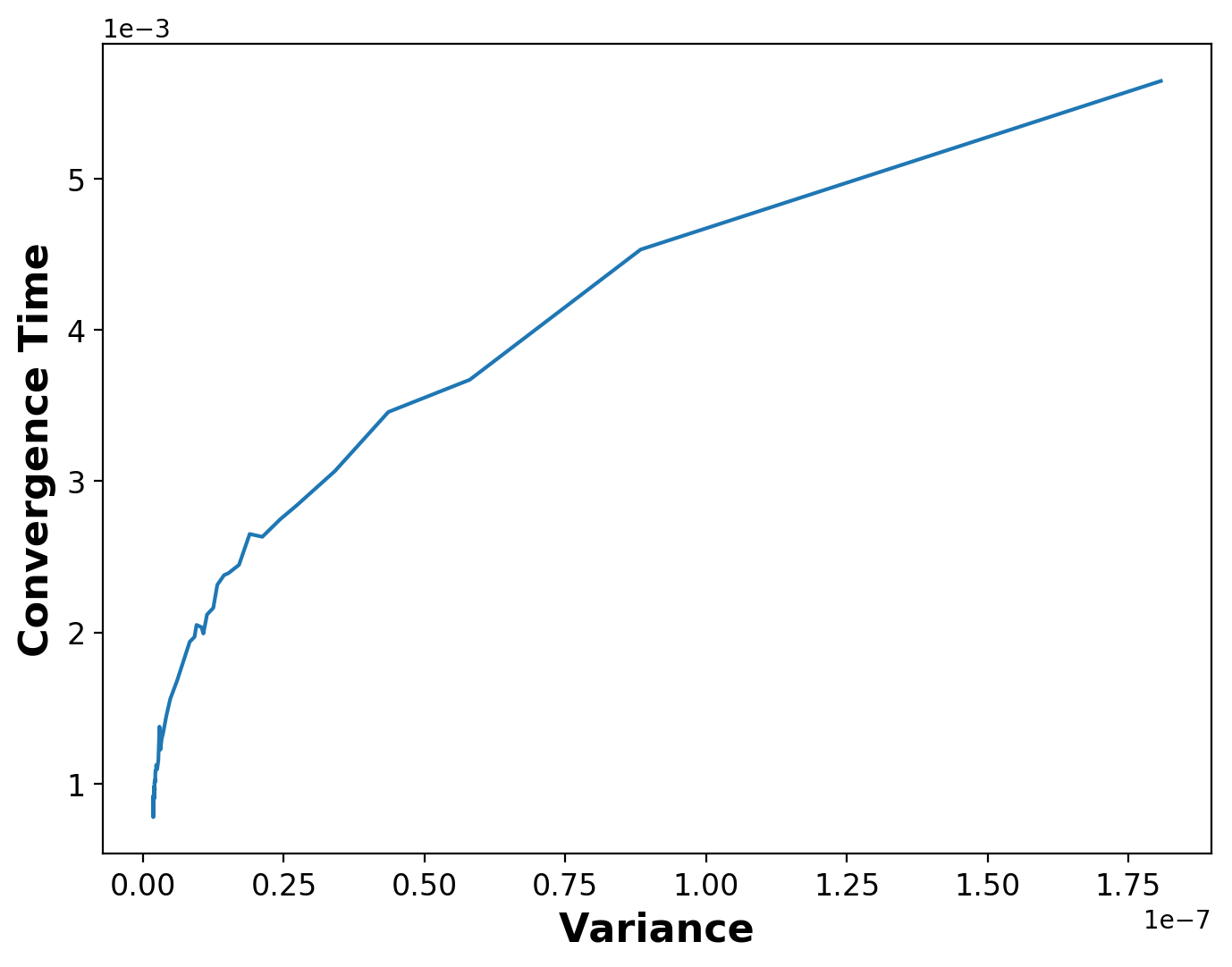}
\includegraphics[scale=0.45]{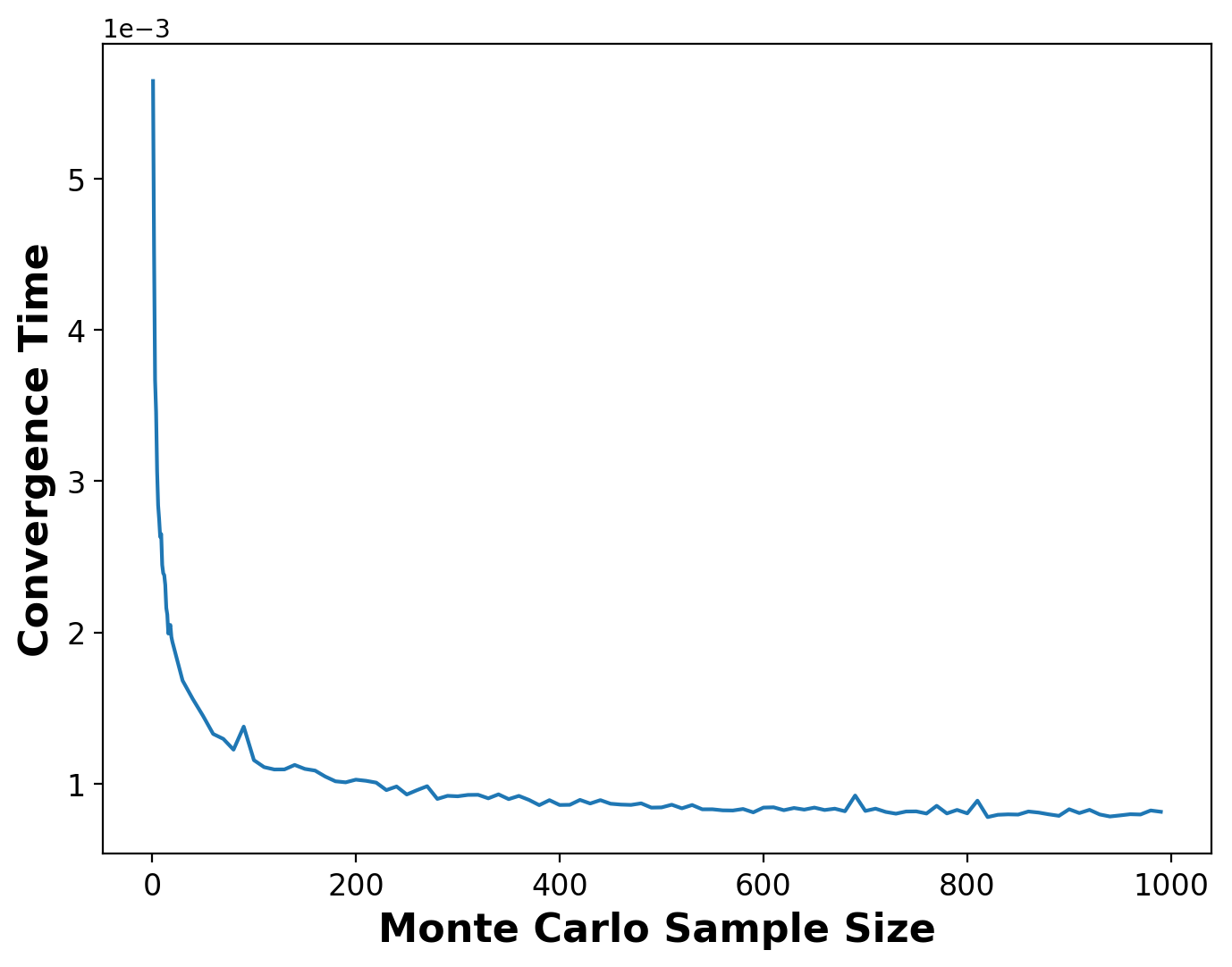}
\caption{Plot of convergence time versus variance of $\widehat{\mathbb{E}}(\tau)$ (left panel) and versus Monte Carlo sample size $N$ (right panel).}
\label{convergence_plot}
\end{figure}
\begin{table}
\centering
\begin{tabular}{|l|l|l|l|l|l|}
\hline
      A-B-C                          & $10$-$10$-$10^5$        & $10^3$-$10$-$10^5$       & $10^5$-$10$-$10^5$  &  $10$-$30$-$10^5$         & $10$-$50$-$10^5$ \\ \hline
time (secs) & 0.4640 & 0.4772   & 0.5152 & 0.3573  & 0.2722  \\ \hline
$\widehat{\mathbb{E}}(\tau)$                          & 0.01  & 0.01  & 0.01 & 0.01  & 0.01 \\ \hline
\end{tabular}
\caption{Results of MC-CAVI for Simulated Example 1.}
\label{my-label}
\end{table}
As with MCEM, $N$ should typically be set to a small number at the beginning of the iterations so that the algorithm can reach fast a region of relatively high probability. $N$ should then be increased   to reduce algorithmic variability close to the convergence region. 
Figure \ref{convergence_plot} shows plots of convergence time versus variance of $\widehat{\mathbb{E}}(\tau)$ (left panel) and versus $N$ (right panel). In VI, iterations are typically terminated when the (absolute) change in the monitored estimate is less than a small threshold. In MC-CAVI the estimate fluctuates around the limiting value after convergence. In the simulation in Figure \ref{convergence_plot}, we terminate the iterations when the difference between the estimated mean (disregarding the first half of the chain) and the true value ($0.01$) is less than $10^{-5}$. Figure \ref{convergence_plot} shows that: (i) convergence time decreases when the variance of $\widehat{\mathbb{E}}(\tau)$ decreases, as anticipated;
 (ii) convergence time decreases when $N$ increases. In (ii), the decrease is most evident when $N$ is still relatively small 
After $N$ exceeds $200$, convergence time remains almost fixed, as the benefit brought by decrease of variance is offset by the cost of extra samples. (This is also in agreement with the policy of $N$ set to a small value at the initial iterations of the algorithm.)

\subsection{Variance Reduction for BBVI}

In non-trivial applications, the variability of the initial estimator $\nabla_{\lambda}\widehat{\textrm{ELBO}}(q)$ within BBVI in (\ref{eq:BBVIest}) will typically be large, so variance reduction approaches such as Rao-Blackwellization and control variates \citep{bbvi} are also used.
Rao-Blackwellization \citep{raoblack} reduces variances by analytically calculating conditional expectations.
In BBVI, within the 
factorization framework of (\ref{eq:meanfield}), where 
$\lambda = (\lambda_1,\ldots, \lambda_b)$, and recalling identity (\ref{eq:mainBB}) for the gradient, 
a Monte Carlo estimator for the gradient with respect to $\lambda_i$, $i\in\{1,\ldots, b\}$, can be simplified as
\begin{equation}
\label{raograd}
\nabla_{\lambda_i}\widehat{\textrm{ELBO}}(q_i) = \tfrac{1}{N}\sum^N_{n=1}\big[\nabla_{\lambda_i}\log q_i(z_i^{(n)}|\lambda_i)\{\log c_i(z_i^{(n)},x)-\log q_i(z_i^{(n)}| \lambda_i)\}\big],
\end{equation}
with $z_i^{(n)} \stackrel{iid}{\sim} q_i(z_i|\lambda_i)$, $1\le n\le N$, and,
\begin{align*} 
c_i(z_i,x):= \exp\big\{\mathbb{E}_{-i}[\log p(z_{i_-},z_{i},z_{i_+},x)]\big\}. 
\end{align*} 
Depending on the model at hand, term $c_i(z_i,x)$ can be obtained analytically  
or via a double Monte Carlo procedure (for estimating $c_i(z_i^{(n)},x)$,
over all $1\le n\le N$) -- or a combination of thereof.
In BBVI, control variates \citep{ross_2002}  can be defined on a per-component basis and be applied to the Rao-Blackwellized noisy gradients of ELBO in (\ref{raograd}) to provide the estimator, 
\begin{equation}
\label{deltaelbo}
\nabla_{\lambda_i}\widehat{\textrm{ELBO}}(q_i) = \tfrac{1}{N}\sum^N_{n=1}\big[\nabla_{\lambda_i}\log q_i(z_i^{(n)}| \lambda_i)\{\log c_i(z_i^{(n)},x)-\log q_i(z_i^{(n)}| \lambda_i)-\widehat{a}^*_i\}\big],
\end{equation}
for the control, 
\begin{equation*}
\widehat{a}^*_i := \frac{\sum^{d_i}_{j=1}\widehat{\textrm{Cov}}(f_{i,j},g_{i,j})}{\sum^{d_i}_{j=1}\widehat{\textrm{Var}}(g_{i,j})},
\end{equation*}
where $f_{i,j}$, $g_{i,j}$ denote the $j$th co-ordinate of the vector-valued functions $f_i$, $g_i$ respectively,
given below,
\begin{align*}
g_i(z_i)&:= \nabla_{\lambda_i}\log q_i(z_i| \lambda_i), \\
f_i(z_i)&:= \nabla_{\lambda_i}\log q_i(z_i| \lambda_i)\{\log c_i(z_i,x)-\log q_i(z_i| \lambda_i)\}.
\end{align*}

\subsection{Simulated Example 2: Model with Hard Constraints}
In this section, we discuss the performance and challenges of MC-CAVI, MCMC, BBVI for models where the support of the posterior -- thus, also the variational distribution -- 
involves hard constraints. 

Here, we provide an example which offers a  simplified version of the NMR problem discussed in Section~\ref{sec:nmr} but allows for the implementation of BBVI, as the involved normalising constants can be easily computed. Moreover, as with other gradient-based methods, BBVI requires to tune the step-size sequence $\{\rho_k\}$ in (\ref{eq:BBVIit}), which might be a laborious task, in particular for increasing dimension. Although there are several proposals aimed to optimise the choice of $\{\rho_k\}$ (\citealp{Bottou2012,advi}), MC-CAVI does not face such a tuning requirement.

We simulate data according to the following scheme: observations 
$\{y_j\}$ are generated from $\mathrm{N}(\vartheta + \kappa_j,\theta^{-1})$, $j = 1,\ldots,n$, with $\vartheta = 6$, $\kappa_j = 1.5\cdot \sin(-2\pi+4\pi(j-1)/{n})$, $\theta = 3$, $n = 100$.  
We fit the following model:
\begin{align*}
\textrm{\underline{Example Model 2}} \\
y_j \mid \vartheta, \kappa_j, \theta&\sim \mathrm{N}(\vartheta + \kappa_j,\theta^{-1}),  \\[0.1cm]
\vartheta &\sim \mathrm{N}(0,10),\\[0.1cm]
\kappa_j  \mid \psi_j &\sim \mathrm{TN}(0,10,-\psi_j,\psi_j),\\
\psi_j \hspace{0.1cm} &\!\!\stackrel{i.i.d.}{\sim} \mathrm{TN}(0.05,10,0,2),\quad j = 1,\ldots,n, \\[0.1cm]
\theta &\sim \mathrm{Gamma}(1,1). 
\end{align*}

\subsubsection*{MCMC}
\label{sec:MCMC}
We use a standard Metropolis-within-Gibbs. We set $y = (y_1, \ldots, y_{n})$, $\kappa = (\kappa_1, \ldots, \kappa_{n})$ and $\psi = (\psi_1, \ldots, \psi_{n})$.
Notice that we have the full conditional distributions, 
\begin{align*}
p(\vartheta| y,\theta, \kappa,  \psi) &= \mathrm{N}\big(\tfrac{\sum^{n}_{j=1}(y_j-\kappa_j)\theta}{\frac{1}{10}+{n}\theta},\tfrac{1}{\frac{1}{10}+{n}\theta}\big),\\[0.1cm] 
p(\kappa_j| y,\theta,\vartheta,  \psi)&= \mathrm{TN}\big(\tfrac{(y_j-\vartheta)\theta}{\frac{1}{10}+\theta},\tfrac{1}{\frac{1}{10}+\theta},-\psi_j,\psi_j\big) ,\\[0.1cm]
p(\theta|y,\vartheta, \kappa,  \psi) &=  \mathrm{Gamma}\big(1+\tfrac{n}{2},1+\tfrac{\sum^{n}_{j=1}(y_j-\vartheta-\kappa_j)^2}{2}\big).
\end{align*}
(Above, and in similar expressions written in the sequel, equality is meant to be properly understood as stating that `the density 
on the left is equal to the density of the distribution on the right'.)
For each $\psi_j$, $1\le j\le {n}$, the full conditional is,
\begin{equation*}
p(\psi_j | y,\theta,\vartheta, \kappa) \propto \frac{ \phi(\tfrac{\psi_j-\frac{1}{20}}{\sqrt{10}})}{\Phi(\tfrac{\psi_j}{\sqrt{10}})-\Phi(\tfrac{-\psi_j}{\sqrt{10}})}\,
\mathbb{I}\,[\,|\kappa_j|<\psi_j<2\,],\quad j = 1,\ldots,{n},
\end{equation*}
where $\phi(\cdot)$ is the density of $\mathrm{N}(0,1)$ and $\Phi(\cdot)$ its cdf.
%
The Metropolis-Hastings proposal for $\psi_j$ is a uniform variate from 
$\textrm{U}(0,2)$.

\subsubsection*{MC-CAVI}

For MC-CAVI, the logarithm of the joint distribution is given by,
\begin{align*}
\log p(y,\vartheta, \kappa, \psi,\theta) &= const. + \tfrac{n}{2}\log \theta - \tfrac{\theta\sum^{n}_{j=1}(y_j - \vartheta-\kappa_j)^2}{2} - \tfrac{\vartheta^2}{2\cdot 10}
-\theta-\sum^{n}_{j=1}\tfrac{\kappa_j^2+(\psi_j-\frac{1}{20})^2}{2\cdot 10}
\\[-0.4cm]
&\qquad \qquad \qquad -\sum^{n}_{j=1} \log(\Phi(\tfrac{\psi_j }{\sqrt{10}})-\Phi(\tfrac{-\psi_j }{\sqrt{10}})),
\end{align*}
under the constraints,
\begin{align*}
 |\kappa_j|<\psi_j<2, \quad j = 1,\ldots,{n}.
\end{align*}
To comply with the above constraints, we factorise the variational distribution as, 
\begin{align}
\label{eq:parts}
q(\vartheta,\theta, \kappa, \psi)=q(\vartheta)q(\theta)\prod^{n}_{j=1}q(\kappa_j,\psi_j).
\end{align}
Here, for the relevant iteration $k$, we have,  
\begin{align*}
q_k(\vartheta) &= 
\mathrm{N}\big(\tfrac{\sum^{n}_{j=1}(y_j-\mathbb{E}_{k-1}(\kappa_j))\mathbb{E}_{k-1}(\theta)}{\frac{1}{10}+{n}\mathbb{E}_{k-1}(\theta)},\tfrac{1}{\frac{1}{10}+{n}\mathbb{E}_{k-1}(\theta)}\big),\\[0.2cm]
q_k(\theta) &= 
\mathrm{Gamma}\big(1+\tfrac{n}{2}, 1+\tfrac{\sum^{n}_{j=1}\mathbb{E}_{k,k-1}((y_j-\vartheta-\kappa_j)^2)}{2})\big), \\[0.3cm]
q_k(\kappa_j,\psi_j) &\propto \exp\big\{- 
\tfrac{\mathbb{E}_{k}(\theta) (\kappa_j-(y_j-\mathbb{E}_{k}(\vartheta)))^2}{2} -\tfrac{\kappa_j^2+(\psi_j-\frac{1}{20})^2}{2\cdot 10} \big\} \big/
 \big(\Phi(\tfrac{\psi_j }{\sqrt{10}})-\Phi(\tfrac{-\psi_j }{\sqrt{10}})\big)\\ &\qquad \qquad\qquad\qquad\qquad\qquad
 \qquad\cdot \mathbb{I}\,[\,|\kappa_j|<\psi_j<2\,],\qquad 1\le j\le {n}.
\end{align*}
The quantity $\mathbb{E}_{k,k-1}((y_j-\vartheta-\kappa_j)^2)$ used in the second line above means that the expectation is considered under $\vartheta\sim q_k(\vartheta)$ and (independently) $\kappa_{j}\sim q_{k-1}(\kappa_{j},\psi_j)$.

Then, MC-CAVI develops as follows:

\begin{itemize}
\item Step 0:  For $k=0$, initialize 
$\mathbb{E}_{0}(\theta)=1$, $\mathbb{E}_{0}(\vartheta)=4$, $\mathbb{E}_{0}(\vartheta^2)=17$.
\item Step $k$: 
 For $k\ge 1$, given $\mathbb{E}_{k-1}(\theta)$, $\mathbb{E}_{k-1}(\vartheta)$,
 execute: 
 \begin{itemize}
 \item For $j=1,\ldots, {n}$, apply an MCMC algorithm -- with invariant law 
 $q_{k-1}(\kappa_j,\psi_j)$ -- consisted of a number, $N$,  of  Metropolis-within-Gibbs iterations carried out over the relevant full conditionals,
\begin{align*}
q_{k-1}(\psi_j| \kappa_j) &\propto\frac{\phi(\tfrac{\psi_j-\frac{1}{20}}{\sqrt{10}})}{\Phi(\tfrac{\psi_j}{\sqrt{10}})-\Phi(\tfrac{-\psi_j}{\sqrt{10}})}\,
\mathbb{I}\,[\,|\kappa_j|<\psi_j<2\,], \\[0.3cm]
q_{k-1}(\kappa_j|\psi_j)&= \mathrm{TN}\big(\tfrac{(y_j-\mathbb{E}_{k-1}(\vartheta))\mathbb{E}_{k-1}(\theta)}{\frac{1}{10}+\mathbb{E}_{k-1}(\theta)},\tfrac{1}{\frac{1}{10}+\mathbb{E}_{k-1}(\theta)},-\psi_j,\psi_j\big).
\end{align*}
As with the full conditional $p(\psi_j | y,\theta,\vartheta,\kappa)$ 
within the MCMC sampler, we use a uniform proposal $\mathrm{U}(0,2)$
at the Metropolis-Hastings step applied for $q_{k-1}(\psi_j| \kappa_j)$. For each $k$, the $N$ iterations begin from the $(\kappa_j,\psi_j)$-values obtained at the end of the corresponding MCMC iterations at step $k-1$,  with very first initial values being $\kappa, \psi_j)=(0,1)$.
Use the $N$ samples to obtain $\mathbb{E}_{k-1}(\kappa_j)$ and $\mathbb{E}_{k-1}(\kappa_j^2)$.
\item Update the variational distribution for $\vartheta$,
\begin{align*}
q_{k}(\vartheta) &= \mathrm{N}\big(\tfrac{\sum^{n}_{j=i}(y_j-\mathbb{E}_{k-1}(\kappa_j))\mathbb{E}_{k-1}(\theta)}{\frac{1}{10}+{n}\mathbb{E}_{k-1}(\theta)},\tfrac{1}{\frac{1}{10}+{n}\mathbb{E}_{k-1}(\theta)}\big)
\end{align*}
and evaluate $\mathbb{E}_{k}(\vartheta)$, $\mathbb{E}_{k}(\vartheta^2)$.
\item Update the variational distribution for $\theta$,
\begin{align*}
q_{k}(\theta)&= \mathrm{Gamma}\big(1+\tfrac{n}{2},1+\tfrac{\sum^{n}_{j=1}\mathbb{E}_{k,k-1}((y_j-\vartheta-\kappa_j)^2)}{2}\big)
\end{align*}
and evaluate $\mathbb{E}_{k}(\theta)$.
 \end{itemize}
\item Iterate until convergence.
\end{itemize}
%

\subsubsection*{BBVI}
For BBVI we assume a variational distribution $q(\theta,\vartheta, \kappa, \psi\,|\,\boldsymbol{\alpha},\boldsymbol{\gamma})$
that factorises as in the case of CAVI in (\ref{eq:parts}), where 
\begin{align*}
\boldsymbol{\alpha} &= (\alpha_{\vartheta}, \alpha_{\theta}, \alpha_{\kappa_1}, \ldots, \alpha_{\kappa_{n}}, \alpha_{\psi_1}, \ldots, \alpha_{\psi_{n}})\ ,  \\ \boldsymbol{\gamma} &= (\gamma_{\vartheta}, \gamma_{\theta}, \gamma_{\kappa_1}, \ldots, \gamma_{\kappa_{n}}, \gamma_{\psi_1}, \ldots, \gamma_{\psi_{n}})
\end{align*}
 to be the variational parameters.
Individual marginal distributions are chosen to agree -- in type -- with the model priors. In particular, we set, 
\begin{align*}
q(\vartheta) &= \mathrm{N}(\alpha_{\vartheta},\exp(\gamma_{\vartheta})),\\[0.2cm]
q(\theta) &= \mathrm{Gamma}(\exp(\alpha_{\theta}),\exp(\gamma_{\theta})), \\[0.2cm]
q(\kappa_j,\psi_j) &= \mathrm{TN}(\alpha_{\kappa_j},\exp(2\gamma_{\kappa_j}),-\psi_j,\psi_j)\otimes \mathrm{TN}(\alpha_{\psi_j},\exp(2\gamma_{\psi_j}),0,2), \quad 1\leq j \leq {n}.
\end{align*}
%
%
%
It is straightforward to derive the required gradients (see Appendix \ref{sec:gradient} for the analytical expressions).
BBVI is applied using Rao-Blackwellization and control variates for variance reduction. The algorithm is as follows,

\begin{itemize}
\item Step 0:  Set $\eta = 0.5$; initialise $\boldsymbol{\alpha}^0 = 0$, $\boldsymbol{\gamma}^0 = 0$ with the exception $\alpha^0_{\vartheta}=4$.
\item Step $k$: 
 For $k\ge 1$, given $\boldsymbol{\alpha}^{k-1}$ and $\boldsymbol{\gamma}^{k-1}$
 execute:
 \begin{itemize}
 \item Draw $(\vartheta^i, \theta^i, \kappa^i,\psi^i)$, for $1\leq i \leq N$,  from $q_{k-1}(\vartheta)$, $q_{k-1}(\theta)$, $q_{k-1}(\kappa,\psi)$.
 \item With the samples, use (\ref{deltaelbo}) to evaluate:
\begin{align*} 
&\nabla^{k}_{\alpha_{\vartheta}}\widehat{\textrm{ELBO}}(q(\vartheta)),\quad \nabla^{k}_{\gamma_{\vartheta}}\widehat{\textrm{ELBO}}(q(\vartheta)), \\  &\nabla^{k}_{\alpha_{\theta}}\widehat{\textrm{ELBO}}(q(\theta)),\quad  \nabla^{k}_{\gamma_{\theta}}\widehat{\textrm{ELBO}}(q(\theta)), \\ &\nabla^{k}_{\alpha_{\kappa_j}}\widehat{\textrm{ELBO}}(q(\kappa_j,\psi_j)), \quad \nabla^{k}_{\gamma_{\kappa_j}}\widehat{\textrm{ELBO}}(q(\kappa_j,\psi_j)),\quad 1\leq j \leq n, \\ &\nabla^{k}_{\alpha_{\psi_j}}\widehat{\textrm{ELBO}}(q(\kappa_j,\psi_j)),\quad 
\nabla^{k}_{\gamma_{\psi_j}}\widehat{\textrm{ELBO}}(q(\kappa_j,\psi_j)), \quad 1\leq j \leq n.
\end{align*}
(Here, superscript $k$ at the gradient symbol $\nabla$ specifies the BBVI iteration.)
 \item Evaluate $\boldsymbol{\alpha}^{k}$ and $\boldsymbol{\gamma}^{k}$:
 \begin{align*}
  (\boldsymbol{\alpha},\boldsymbol{\gamma})^{k} &= (\boldsymbol{\alpha},\boldsymbol{\gamma})^{k-1} + \rho_k\nabla^{k}_{(\boldsymbol{\alpha},\boldsymbol{\gamma})}\widehat{\textrm{ELBO}}(q),
 \end{align*}
where $q = (q(\vartheta), q(\theta), q(\kappa_1, \psi_1), \ldots, q(\kappa_n, \psi_n))$. For the learning rate, we  employed the AdaGrad algorithm \citep{duchi2011adaptive} and set $\rho_k = \eta \, \textrm{diag}(G_k)^{-1/2}$, where $G_k$ is a matrix equal to the sum of the first $k$ iterations of the outer products of the gradient, and $\textrm{diag}(\cdot)$ maps a matrix to its diagonal version.
 \end{itemize}
\item Iterate until convergence.
\end{itemize}

\subsubsection*{Results}
\begin{figure}[!h]
\begin{center}
\includegraphics[scale=0.35]{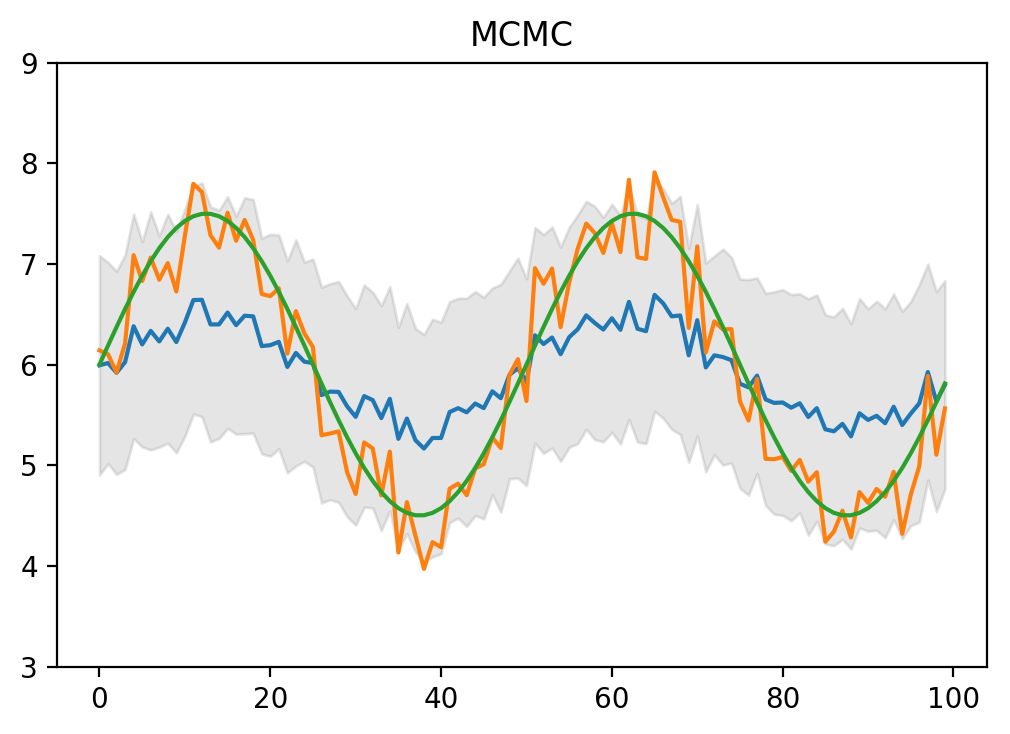}
\includegraphics[scale=0.35]{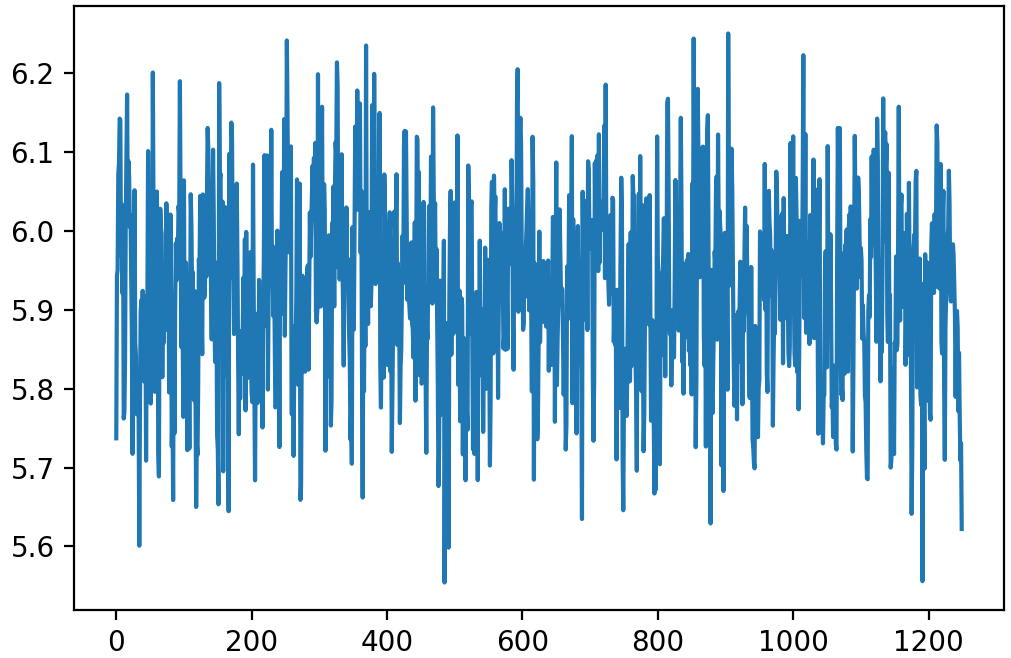}
\includegraphics[scale=0.35]{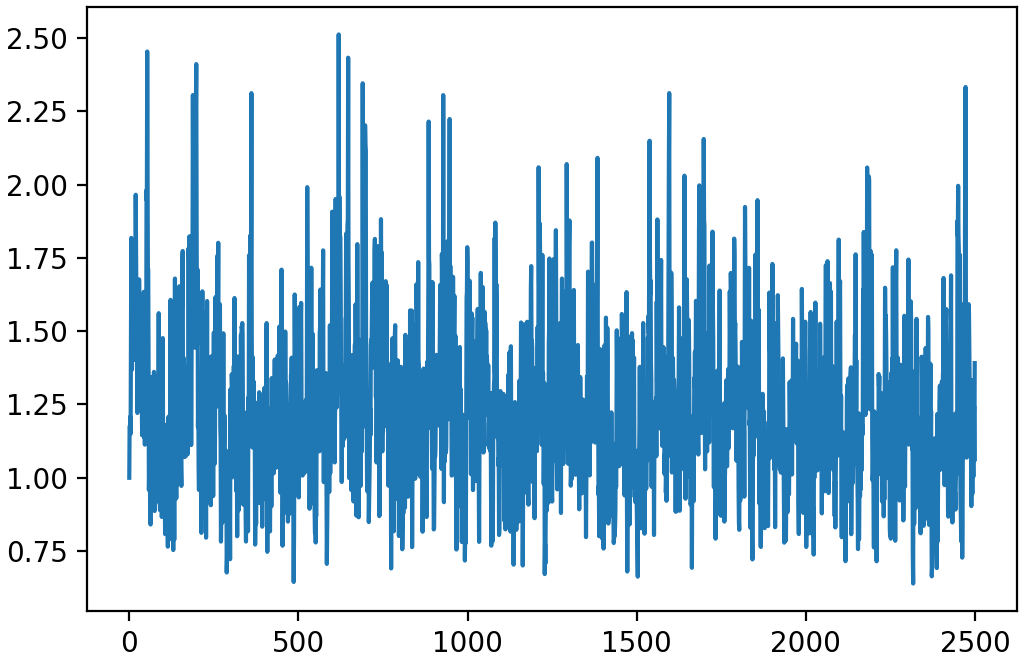}
\includegraphics[scale=0.35]{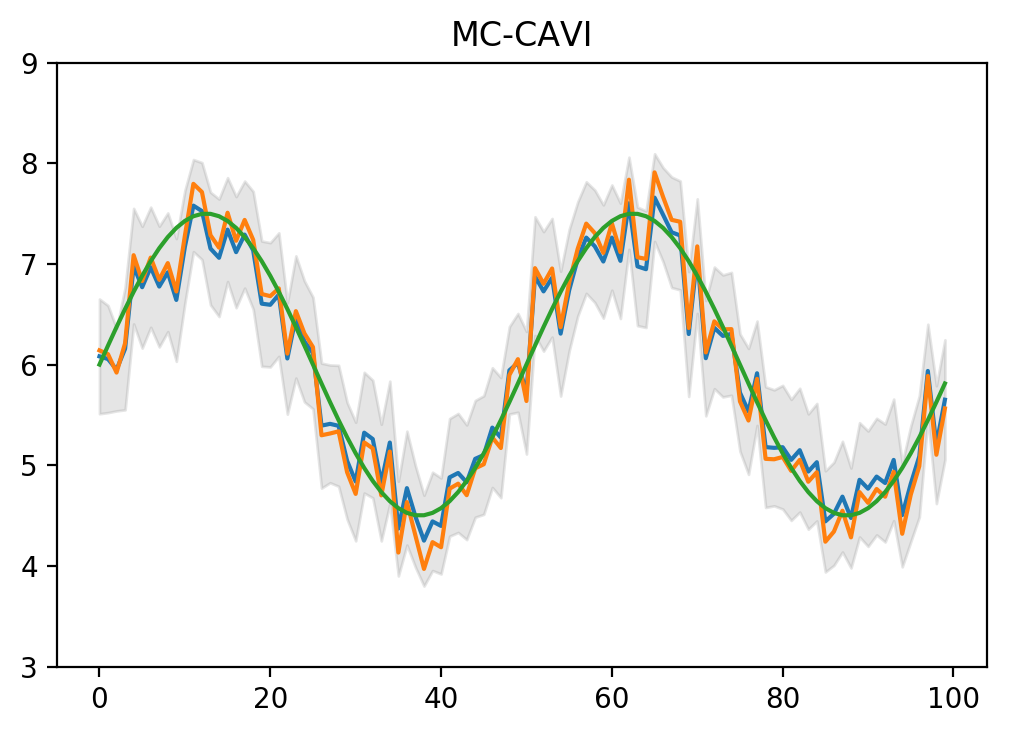}
\includegraphics[scale=0.35]{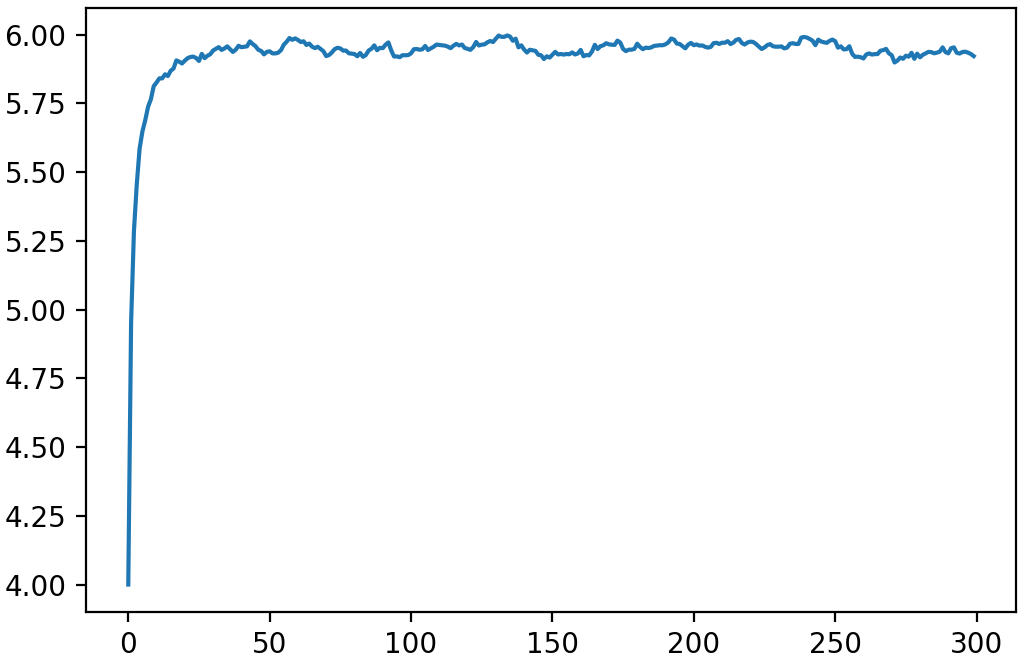}
\includegraphics[scale=0.35]{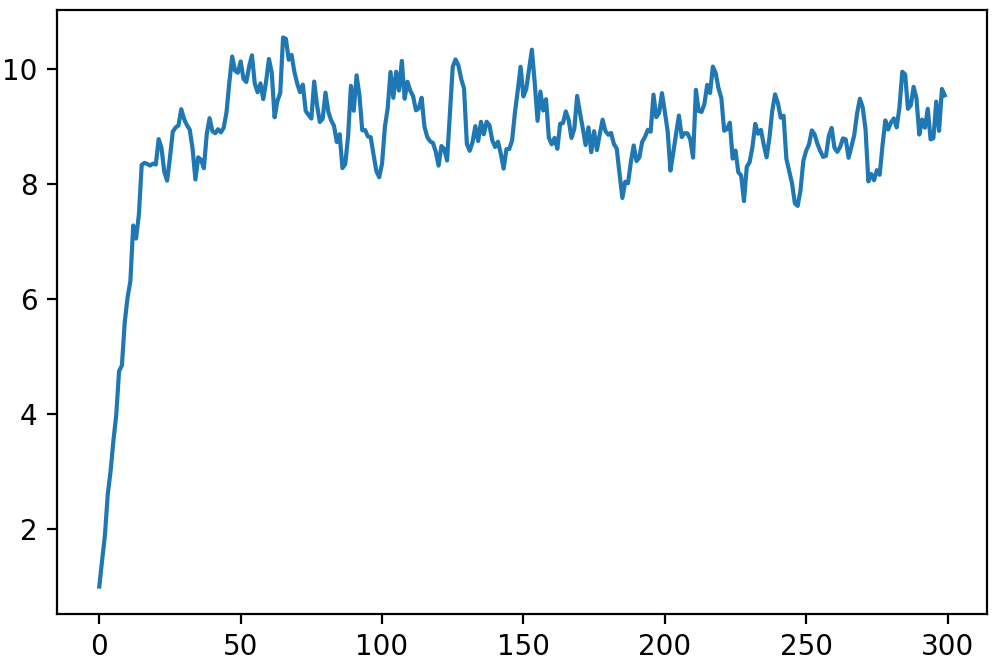}
\includegraphics[scale=0.35]{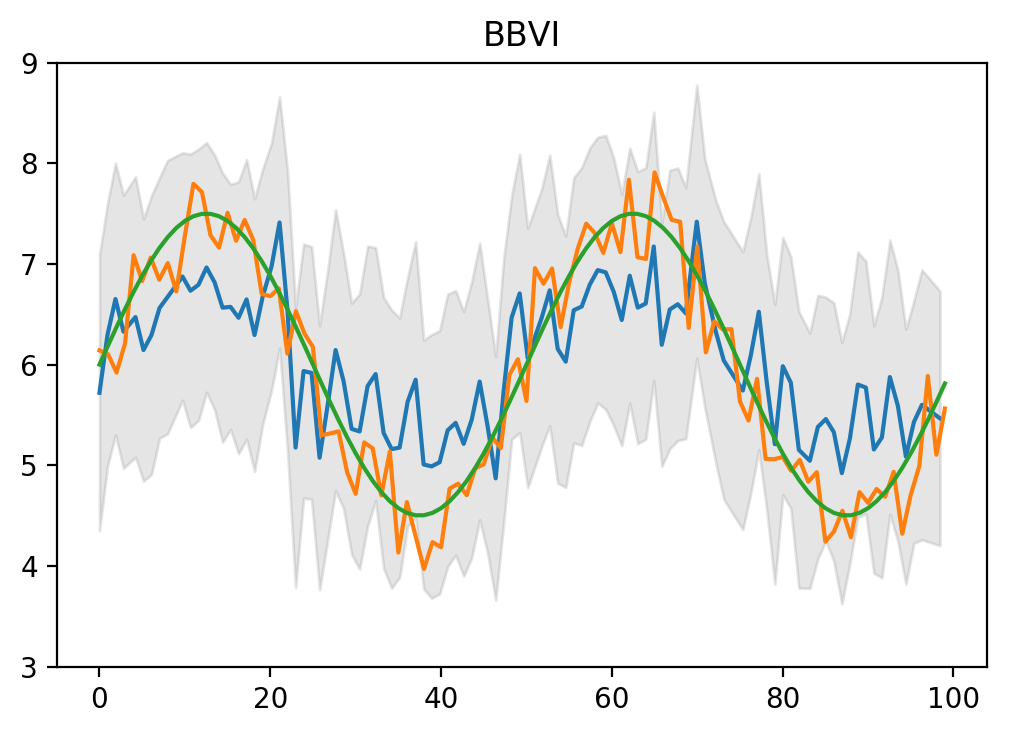}
\includegraphics[scale=0.35]{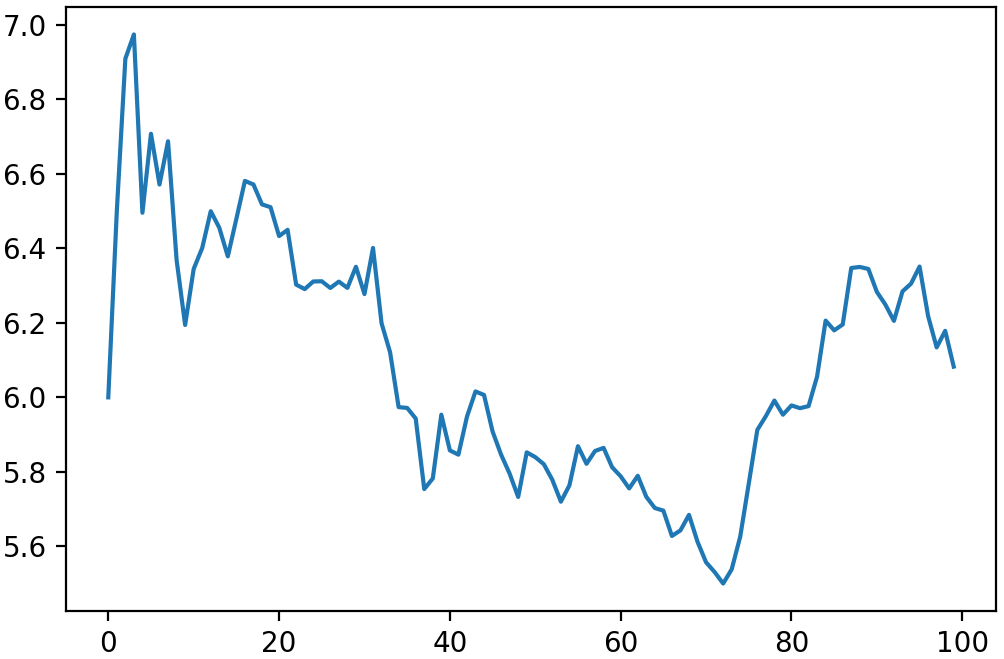}
\includegraphics[scale=0.35]{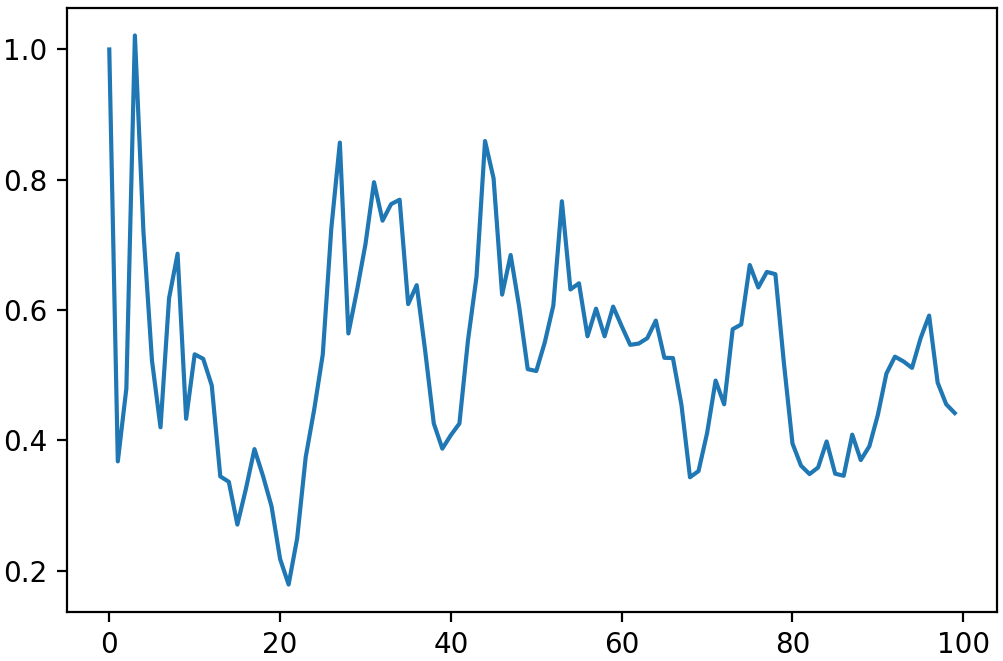}
\end{center}
\vspace{-0.3cm}
\caption{Model fit (left panel), traceplots of $\vartheta$ (middle panel) and traceplots of $\theta$ (right panel) for the three algorithms: MCMC (first row), MC-CAVI (second row) and BBVI (third row) -- for Example Model 2 -- when allowed $100$secs of execution. In the plots showing model fit, the green line represents the data without noise, the orange line the data with noise; the blue line shows the corresponding posterior means and the grey area the pointwise 95\% posterior credible intervals.}
\label{resultplot}
\end{figure}

The  three algorithms have different stopping criteria. We run each for $100$secs for parity. A summary of results is given in Table \ref{resulttable}. Model fitting plots and algorithmic traceplots are shown in Figure \ref{resultplot}.

\begin{table}[!h]
\begin{tabular}{|l|l|l|l|}
\hline
           & MCMC                                                                           & MC-CAVI                                                                                     & BBVI                                                                           \\ \hline
Iterations & \begin{tabular}[c]{@{}l@{}}No. Iterations = 2,500\\ Burn-in = 1,250\end{tabular} & \begin{tabular}[c]{@{}l@{}}No. Iterations = 300\\ $N = 10$\\ Burn-in = 150\end{tabular} & \begin{tabular}[c]{@{}l@{}}No. Iterations = 100\\ $N = 10$\end{tabular} \\ \hline
$\vartheta$         & 5.927 (0.117)                                                                   & 5.951 (0.009)                                                                                & 6.083 (0.476)                                                                   \\ \hline
$\theta$        & 1.248 (0.272)                                                                   & 8.880 (0.515)                                                                                & 0.442 (0.172)                                                                   \\ \hline
\end{tabular}
\caption{Summary of results: last two rows show the average for the corresponding parameter (in horizontal direction) and algorithm (in vertical direction), after burn-in (the number in brackets is the corresponding standard deviation). All algorithms were executed for $10^2$secs. The first row gives some algorithmic details.}
\label{resulttable}
\end{table}

\begin{figure}[!h]
\begin{center}
\includegraphics[scale=0.5]{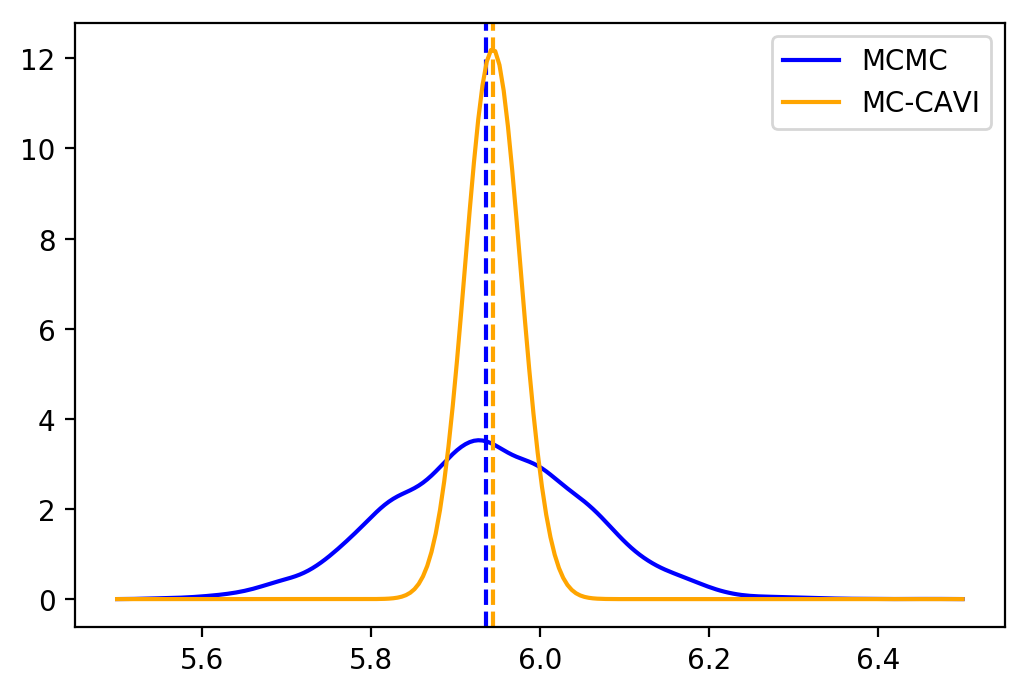}
\end{center}
\vspace{-0.3cm}
\caption{Density plots for the true posterior of $\vartheta$ (blue line) -- obtained via an expensive MCMC --  and the corresponding approximate distribution provided by MC-CAVI.}
\label{resultdensity}
\end{figure}

\noindent Table \ref{resulttable} indicates that all three algorithms approximate the posterior mean of $\vartheta$ effectively; the estimate from MC-CAVI has smaller variability than the one of BBVI; the opposite holds for the variability in the estimates for $\theta$. 
Figure \ref{resultplot} shows that the traceplots for BBVI are unstable, a sign that the gradient estimates have high variability. In contrast, MCMC and MC-CAVI perform rather well. Figure \ref{resultdensity} shows the `true' posterior density of $\vartheta$ (obtained from an expensive MCMC with 10,000 iterations -- 5,000 burn-in) and the corresponding approximation obtained via MC-CAVI. In this case, the variational approximation is quite accurate at the estimation of the mean  but underestimates the posterior variance (rather typically for a VI method). We mention that for BBVI we also tried to use normal laws as variational distributions -- as this is mainly the standard choice in the literature  -- however, in this case, the performance of BBVI deteriorated
even further.

\section{Application to $^1$H NMR Spectroscopy}
\label{sec:nmr}


We demonstrate the utility of MC-CAVI in a statistical model
proposed in the field of metabolomics by \cite{batmanmodel}, and used in NMR (Nuclear Magnetic Resonance) data analysis. 
Proton nuclear magnetic resonance ($^1$H NMR) is an extensively used technique for measuring abundance (concentration) of a number of metabolites in complex biofluids. 
NMR spectra are widely used in metabolomics to obtain profiles of metabolites present in biofluids.  
The NMR spectrum can contain information for a few hundreds of compounds. 
Resonance peaks generated by each compound must be identified in the spectrum after  
deconvolution. The spectral signature of a compound is given by a combination of peaks not necessarily close to each other. Such compounds can generate hundreds of resonance peaks, many of which overlap. This causes difficulty in peak identification and deconvolution. The analysis of NMR spectrum is further complicated by fluctuations in peak positions among spectra induced by uncontrollable variations in experimental conditions and the chemical properties of the biological samples, e.g.~by the pH. 
Nevertheless, extensive information on the patterns of spectral resonance generated by human metabolites is now available in online databases. By incorporating this information into a Bayesian model, we can deconvolve resonance peaks from a spectrum and obtain explicit concentration estimates for the corresponding metabolites. Spectral resonances that cannot be deconvolved in this way may also be of scientific interest; these are modelled in \cite{batmanmodel} using wavelet basis functions. 
More specifically, 
an NMR spectrum is a collection of 
peaks convoluted with various horizontal translations and vertical scalings, 
with each peak having the form of a Lorentzian curve. A number of metabolites of interest 
have known NMR spectrum shape, with the height of the peaks or their width in a particular experiment providing information about the abundance of each metabolite.  

\begin{figure}[!h]
\begin{center}
\includegraphics[scale=0.7]{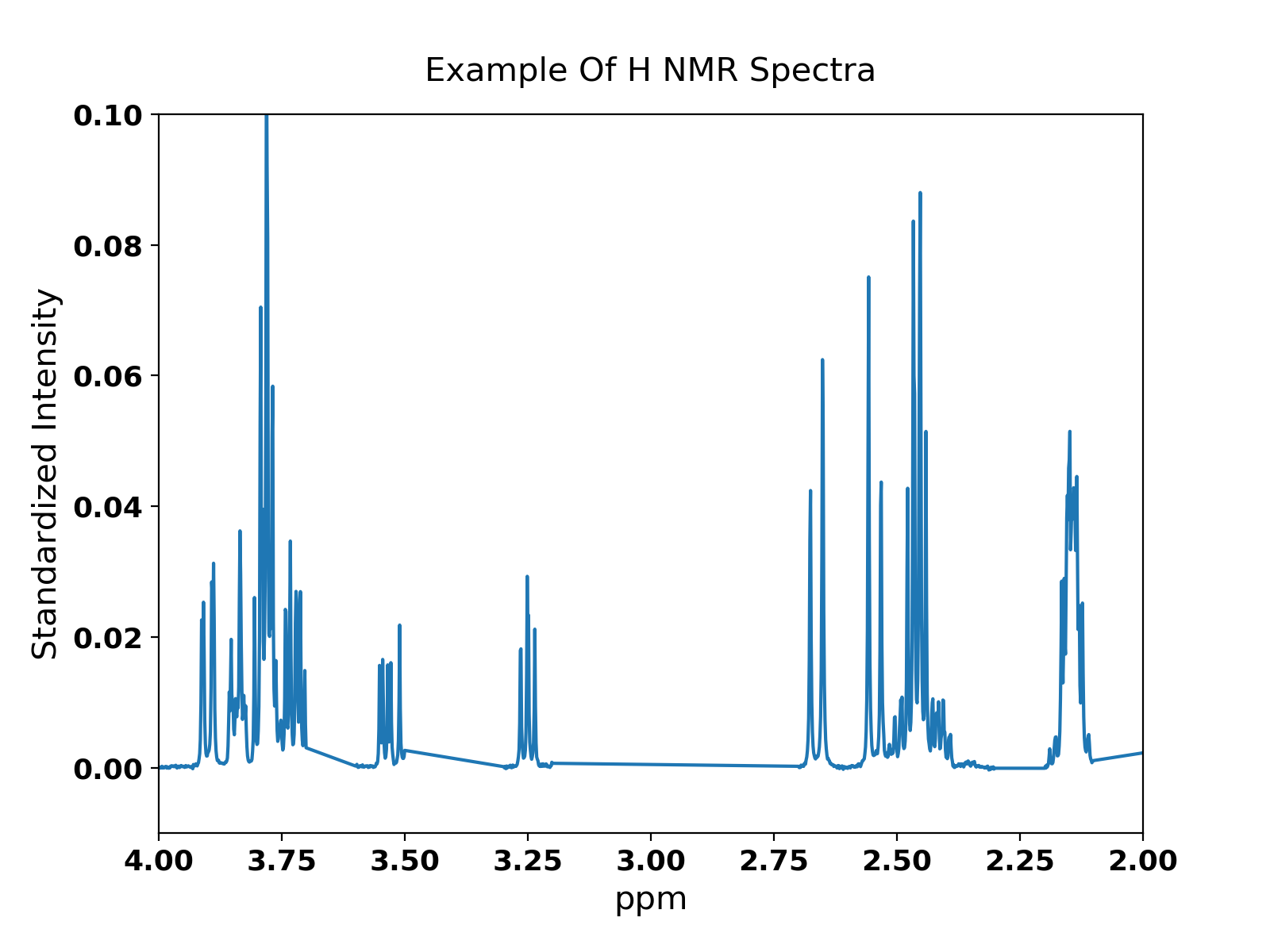}
\end{center}
\vspace{-0.7cm}
\caption{An Example of $^1$H NMR spectrun.}
\label{nmrexample}
\end{figure}

The zero-centred, standardized Lorentzian function is defined as:
\begin{equation}
\ell_\gamma(x) = \frac{2}{\pi}\frac{\gamma}{4x^2+\gamma^2}
\end{equation}
where $\gamma$ is the peak width at half height. 
An example of $^1$H NMR spectrum is shown in Figure \ref{nmrexample}. The x-axis of the spectrum measures chemical shift in parts per million (ppm) and corresponds to the resonance frequency. The y-axis measures relative resonance intensity. 
Each spectrum peak corresponds to magnetic nuclei resonating at a particular frequency in the biological mixture, with every metabolite having a characteristic molecular $^1$H NMR `signature'; the result is a convolution of Lorentzian peaks that appear in specific positions in $^1$H NMR spectra. Each metabolite in the experiment usually gives rise to  more  than a `multiplet' in the spectrum -- 
i.e.~linear combination of Lorentzian functions, symmetric around a central point. 
Spectral signature (i.e.~pattern multiplets) of many metabolites are stored in public databases. 
The aim of the analysis is: (i) to deconvolve resonance peak in the spectrum and assign them to a particular metabolite; (ii) estimate the abundance of the catalogued metabolites; (iii) model  the component of a spectrum that cannot be assigned to known compounds. \cite{batmanmodel}  propose a two-component joint model for a spectrum, in which the metabolites whose peaks we wish to assign explicitly are modelled  parametrically, using information from the online databases, while the unassigned spectrum is modelled using wavelets.

\subsection{The Model}

We now describe the model of \cite{batmanmodel}. The available data are represented by the pair $(\mathbf{x},\mathbf{y})$, where $\mathbf{x}$ is a vector of $n$ ordered points (of the order $10^3-10^4$) on the chemical shift axis -- often regularly spaced -- and $\mathbf{y}$ is the vector of the corresponding resonance intensity measurements (scaled, so that they sum up to $1$). 
The conditional law of $\mathbf{y}|\mathbf{x}$ is modelled under the assumption that $y_i| \mathbf{x}$ are independent normal variables and,
\begin{equation}
\mathbb{E}\,[\,y_i\,|\,\mathbf{x}\,] = \phi(x_i) + \xi(x_i), \quad 1\leq i \leq n.
\end{equation}
Here, the $\phi$ component of the model represents signatures that we wish to assign to target 
metabolites.
The $\xi$ component models signatures of remaining metabolites present in the spectrum, but not explicitly modelled. We refer to this latter as residual spectrum and we highlight the fact that it is important to account for it as it can unveil important information not captured by $\phi(\cdot)$. Function $\phi$ is constructed parametrically using results from the physical theory of NMR and information available online databases or expert knowledge, while $\xi$ is modelled semiparametrically with wavelets generated by a mother wavelet (symlet 6) that resembles the Lorentzian curve. 

More analytically, 
\begin{equation*}
\phi(x_i) = \sum_{m=1}^{M}t_m(x_i)\beta_{m}
\end{equation*} 
where $M$ is the number of metabolites modelled explicitly and  $\beta = (\beta_{1},\ldots,\beta_{M})^{\top}$ is a parameter vector corresponding to metabolite concentrations. 
Function  $t_m(\cdot)$ represents a continuous template function that specifies the NMR signature of metabolite $m$ and it is defined as,
\begin{equation}
t_m(\delta) = \sum_u \sum^{V_{m,u}}_{v=1}z_{m,u}\,\omega_{m,u,v}\,\ell_{\gamma}(\delta-\delta^*_{m,u}-c_{m,u,v}),\quad \delta>0,
\end{equation}
where $u$ is an index running over all multiplets  assigned to metabolite $m$, 
 $v$ is an index representing a peak in a multiplet and  $V_{m,u}$ is the number of peaks in multiplet $u$ of metabolite $m$. In addition, 
$\delta^*_{m,u}$ specifies the theoretical position on the chemical shift axis of the centre of mass of the $u$th multiplet of the $m$th metabolite; $z_{m,u}$ is a positive quantity, usually equal to the number of protons in a molecule of
metabolite $m$ that contributes to the resonance signal of multiplet $u$; $\omega_{m,u,v}$ is the weight determining the relative heights of the peaks of the multiplet; $c_{m,u,v}$ is the translation determining the horizontal offsets of the peaks from the centre of mass of the multiplet. Both $\omega_{m,u,v}$ and $c_{m,u,v}$ can be computed by empirical estimates of the so-called $J$-coupling constants; see \cite{hore2015nuclear} for more details. The $z_{m,u}$'s and $J$-coupling constants information can be found in online databases or from expert knowledge. 

The residual spectrum is modelled through wavelets,

\begin{equation*}
\xi(x_i) = \sum_{j,k}\varphi_{j,k}(x_i)\vartheta_{j,k}
\end{equation*}
where 
$\varphi_{j,k}(\cdot)$
 denote the orthogonal wavelet functions generated by the symlet-6 mother wavelet, see \cite{batmanmodel} for full details; here,  
$\vartheta = (\vartheta_{1,1},\ldots,\vartheta_{j,k},\ldots)^{\top}$ is the vector of wavelet coefficients. Indices $j,k$ correspond to the $k$th wavelet in the $j$th scaling level.

Finally, overall, the model for an NMR spectrum can be re-written in matrix form as: 
\begin{equation}
\mathcal{W}(\mathbf{y} -\mathbf{T} \beta) = \mathbf{I}_{n_1} \vartheta  + \epsilon, \quad \boldsymbol{\epsilon} \sim \mathrm{N}(0,\mathbf{I}_{n_1}/\theta),
\label{nmrlikelihood}
\end{equation}
where $\mathcal{W}\in \mathbb{R}^{n\times {n_1}}$ is the inverse wavelet transform, 
$M$ is the total number of known metabolites,
$\mathbf{T}$ is an $n \times M$ matrix with its $(i,m)$th entry equal to $t_m(x_i)$
and  $\theta$ is a scalar precision parameter.

\subsection{Prior Specification} 
\label{sec:priordist}
\cite{batmanmodel} assign the following prior distribution to the parameters in the Bayesian model.
For the concentration parameters, we assume 
\begin{equation*}
\beta_m \sim \mathrm{TN}(e_m,1/s_m,0,\infty),
\end{equation*}
where $e_m = 0$ and $s_m = 10^{-3}$, for all $m=1,\ldots, M$. Moreover,
\begin{align*}
\gamma &\sim \mathrm{LN}(0,1); \\
%
\delta^*_{m,u} &\sim \mathrm{TN}(\hat{\delta}^*_{m,u},10^{-4},\hat{\delta}^*_{m,u}-0.03,\hat{\delta}^*_{m,u}+0.03),
\end{align*}
where LN denotes a log-normal distribution and $\hat{\delta}^*_{m,u}$ is the estimate for $\delta^*_{m,u}$ obtained from the online database HMDB \citep[see][]{hmdb1, hmdb2, hmdb3, hmdb4}. In the regions of the spectrum where both parametric (i.e.~$\phi$) and semiparametric (i.e.~$\xi$) components  need to be fitted, the likelihood is unidentifiable. To tackle this problem, \cite{batmanmodel} opt for shrinkage priors for the wavelet coefficients and include a vector of hyperparameters $\psi$ -- each component $\psi_{j,k}$ of which corresponds to a wavelet coefficient -- to penalize the semiparametric component. To reflect prior knowledge that NMR spectra are usually restricted to the half plane above the chemical shift axis, \cite{batmanmodel} introduce a vector of hyperparameters $\tau$, each component of which, $\tau_i$, corresponds to a spectral data point, to further penalize spectral reconstructions in which some components of $\mathcal{W}^{-1}\boldsymbol{\vartheta}$ are less than a small negative threshold. In conclusion, \cite{batmanmodel} specify the following joint prior density for $(\vartheta, 
\psi,\tau,\theta)$,
\begin{align*}
p(\vartheta,  \psi, \tau,\theta) &\propto \theta^{a+\tfrac{n+n_1}{2}-1} \Big\{\prod_{j,k}\psi^{c_j-0.5}_{j,k}\exp\big(-\tfrac{\psi_{j,k} d_j}{2}\big)\Big\}\\
&\qquad \qquad \times \exp\Big\{-\tfrac{\theta}{2}\Big(e+\sum_{j,k}\psi_{j,k}\,\vartheta^2_{j,k}+ r\sum^{n}_{i=1}(\tau_i-h )^2\Big)\Big\} \\ 
&\qquad \qquad \qquad \qquad \times \mathbbm{1}\,\big\{\,\mathcal{W}^{-1} \vartheta\geq \tau,\,\, h\mathbf{1}_{n}\geq \tau\,\big\},
%
%
%
%
\end{align*}
where $ \psi$ introduces local shrinkage for the marginal prior of $\vartheta$ and $\tau$ is a vector of $n$ truncation limits, which bounds $\mathcal{W}^{-1} \vartheta$ from below. The truncation imposes an identifiability constraint: without it, when the signature template does not match the shape of the spectral data, the mismatch will be compensated by negative wavelet coefficients, such that an ideal overall model fit is achieved even though the signature template is erroneously assigned and the concentration of metabolites is overestimated. Finally we set  $c_j = 0.05$, $d_j = 10^{-8}$, $h = -0.002$, $r = 10^5$, $a = 10^{-9}$, $e = 10^{-6}$;  see \cite{batmanmodel} for more details.

\subsection{Results}

BATMAN is an $\mathsf{R}$ package for estimating metabolite concentrations from NMR spectral data using a specifically designed  MCMC algorithm \citep{batman} to perform posterior inference from the Bayesian model described above. 
We implement a MC-CAVI version of BATMAN 
and compare its performance with the original MCMC algorithm.
Details of the implementation of MC-CAVI are given in Appendix \ref{sec:BATMAN}.
Due to the complexity of the model and the datasize, it is challenging for both algorithms to reach convergence. We run the two methods, MC-CAVI and MCMC, for approximately an equal amount of time, to analyse a full spectrum with 1,530 data points and modelling parametrically 10 metabolites. We fix the number of iterations for MC-CAVI to 1,000, with a burn-in of 500 iterations; 
we set the Monte Carlo size to $N=10$ for all iterations. 
The execution time for this MC-CAVI algorithms is $2,048$secs.
For the MCMC algorithm, we fix the number of iterations to 2,000, with a burn-in of 1,000 iterations. This MCMC algorithm has an execution time of $2,098$secs.

\begin{figure}[!h]
\begin{center}
\includegraphics[scale=0.4]{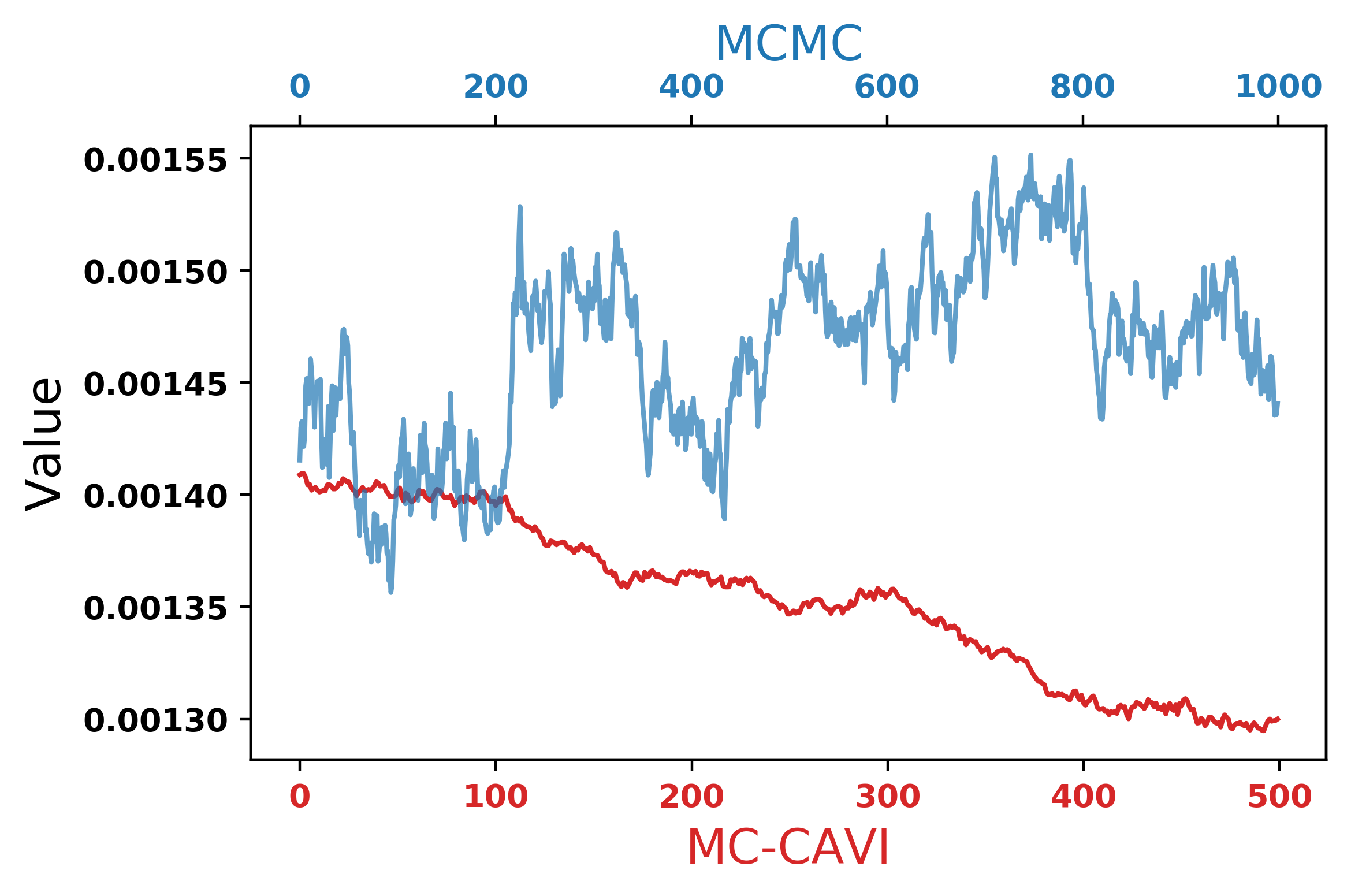}
\includegraphics[scale=0.4]{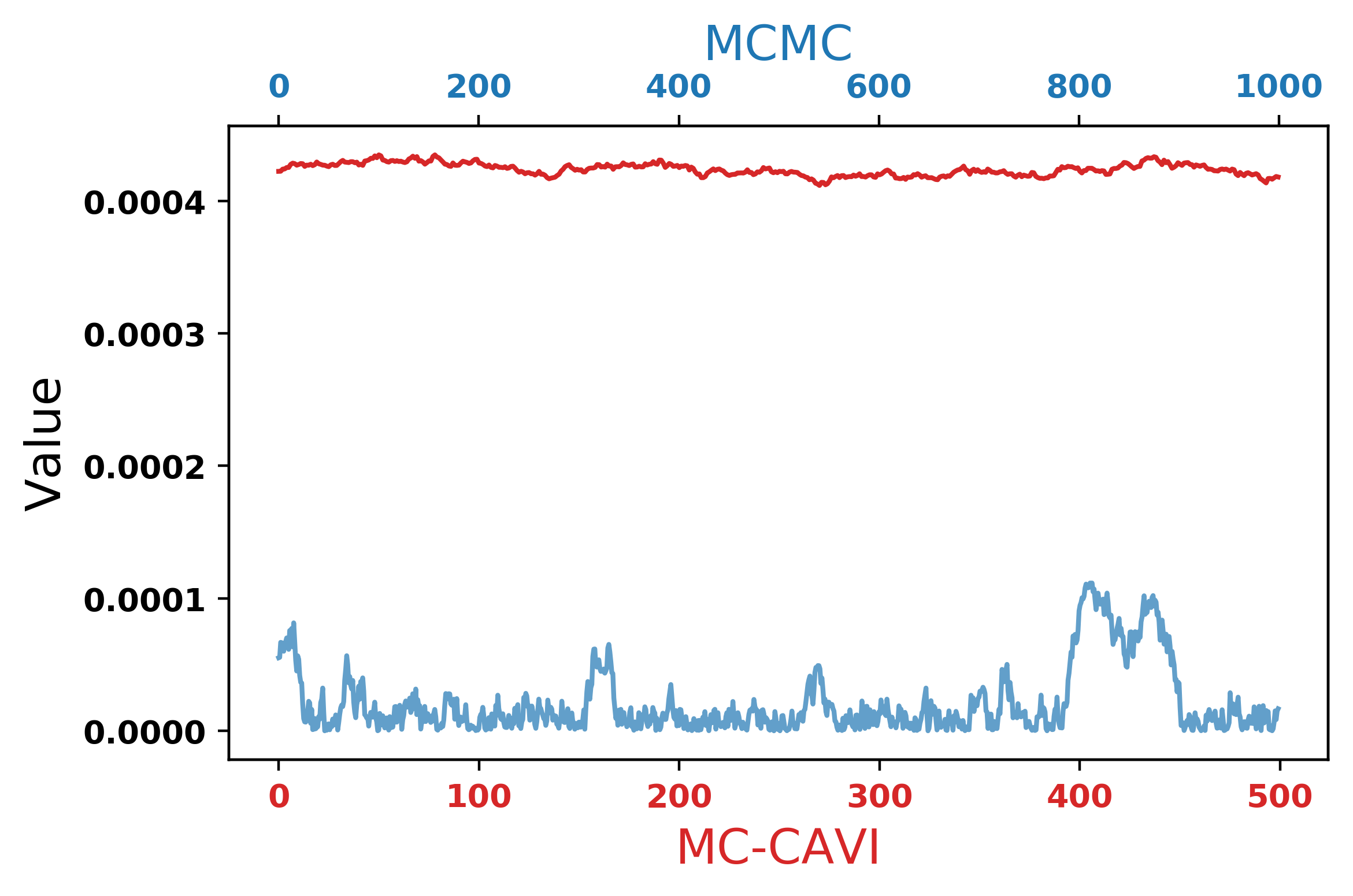}
\includegraphics[scale=0.4]{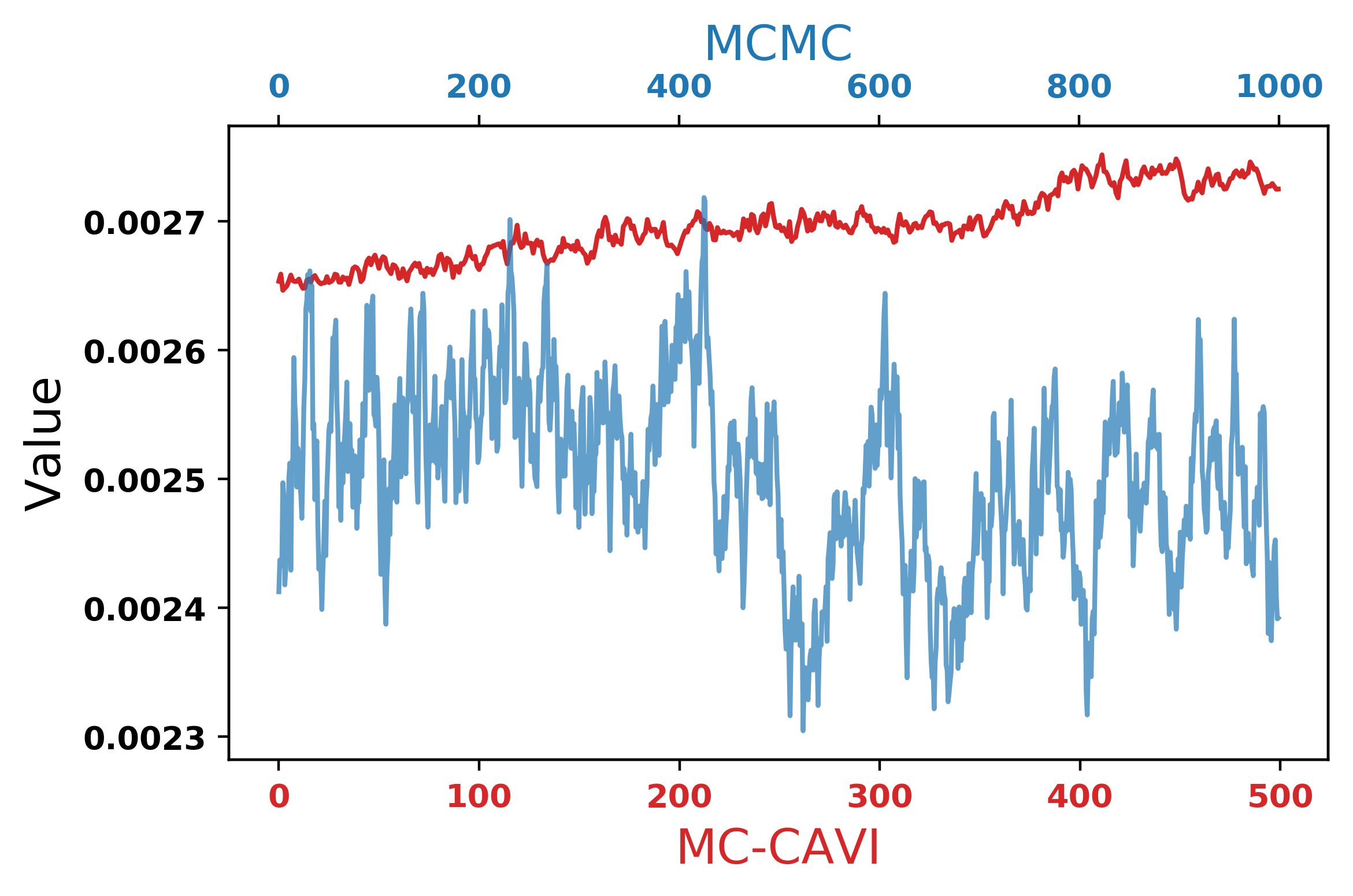}
\includegraphics[scale=0.4]{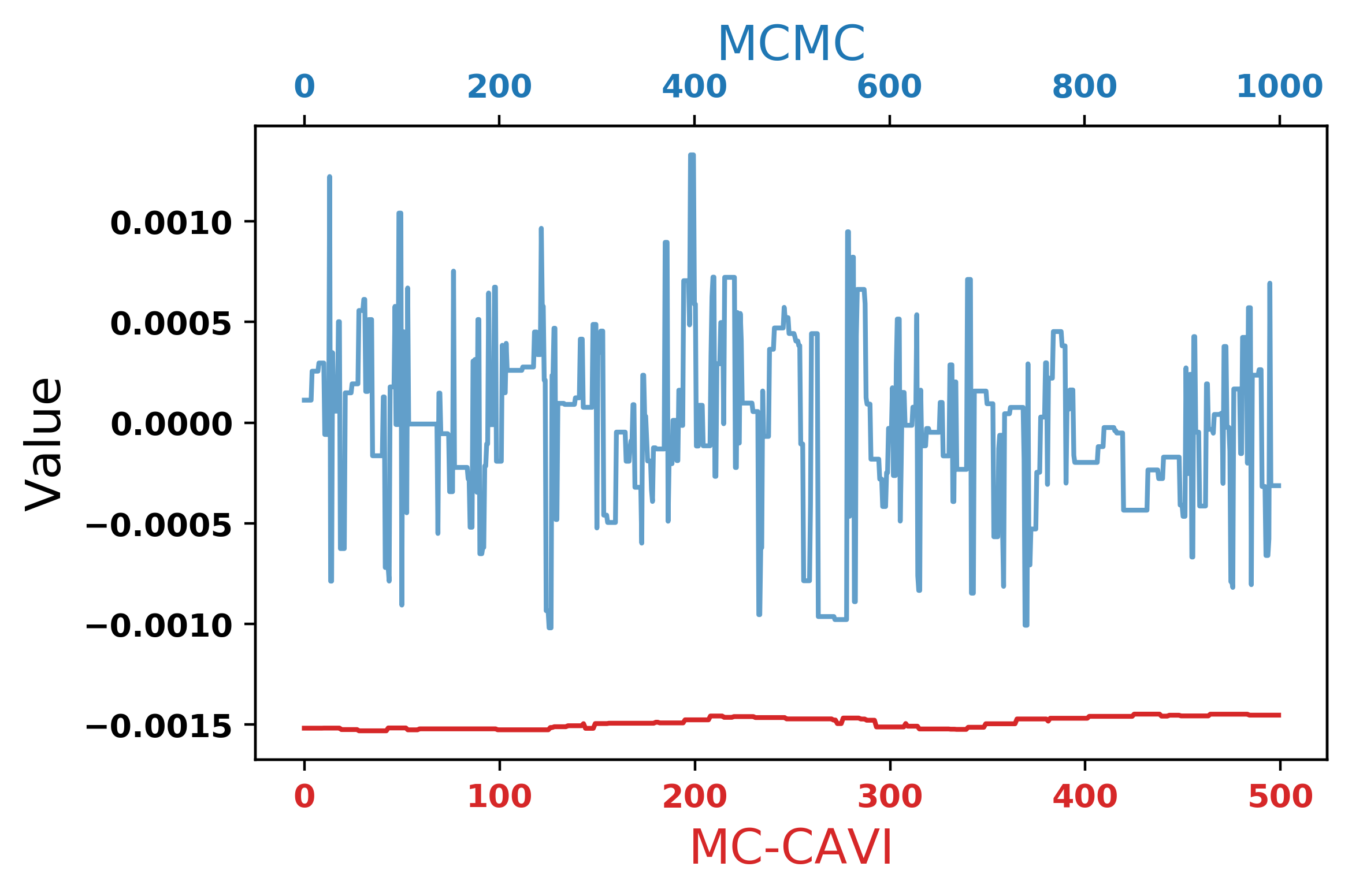}
\end{center}
\vspace{-0.3cm}
\caption{Traceplots of Parameter Value against Number of Iterations after the burn-in period for $\beta_3$ (upper left panel), $\beta_4$ (upper right panel), $\beta_9$ (lower left panel) and $\delta_{4,1}$ (lower right panel). The $y$-axis corresponds to the obtained parameter values (the mean of the distribution $q$ for MC-CAVI and traceplots for MCMC). The red line shows the results from MC-CAVI and the blue line from MCMC. Both algorithms are executed for the same (approximately) amount of time.}
\label{paracomparison}
\end{figure}

\begin{figure}[!h]
\begin{center}
\includegraphics[scale=0.5]{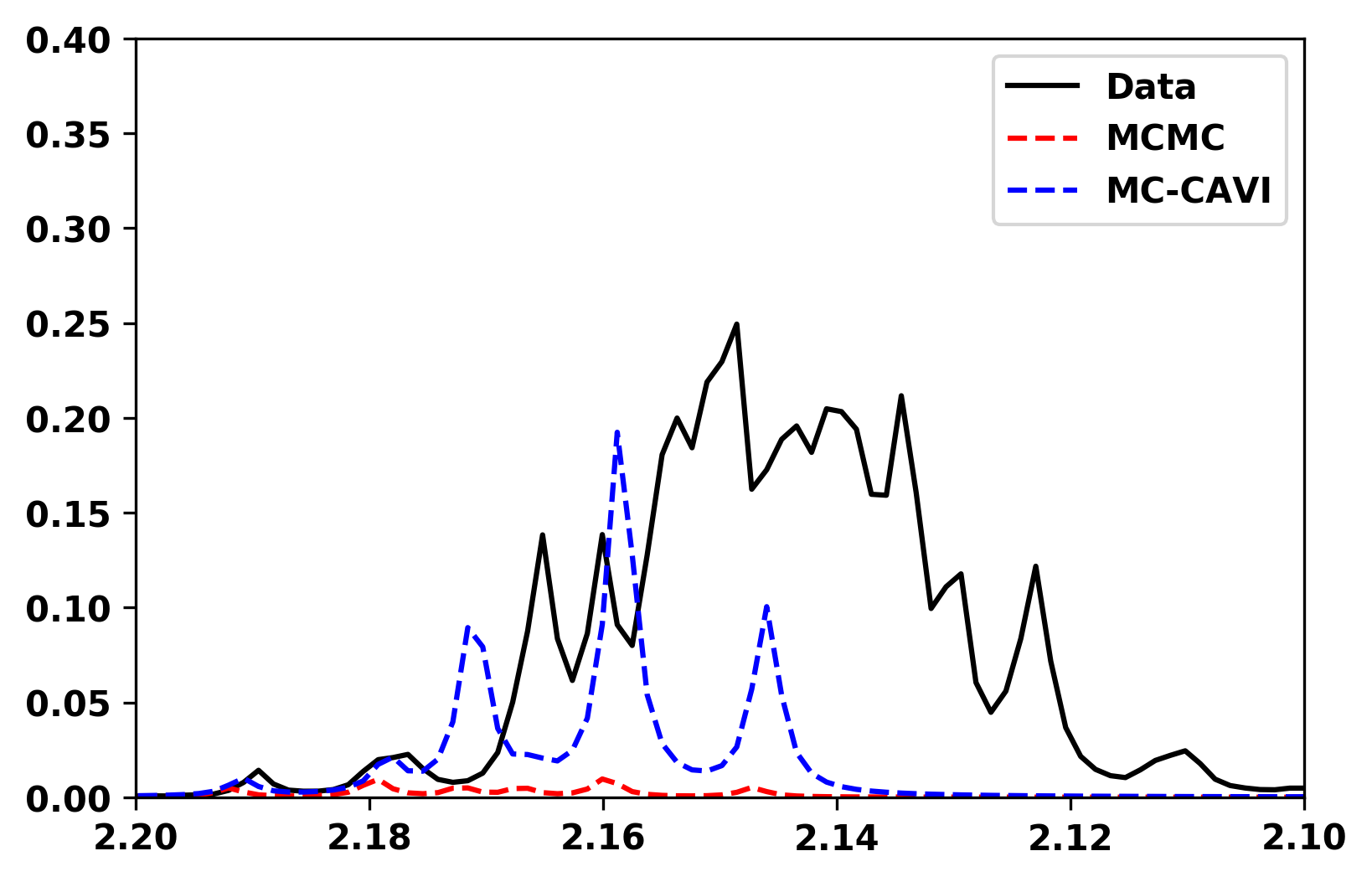}
\includegraphics[scale=0.5]{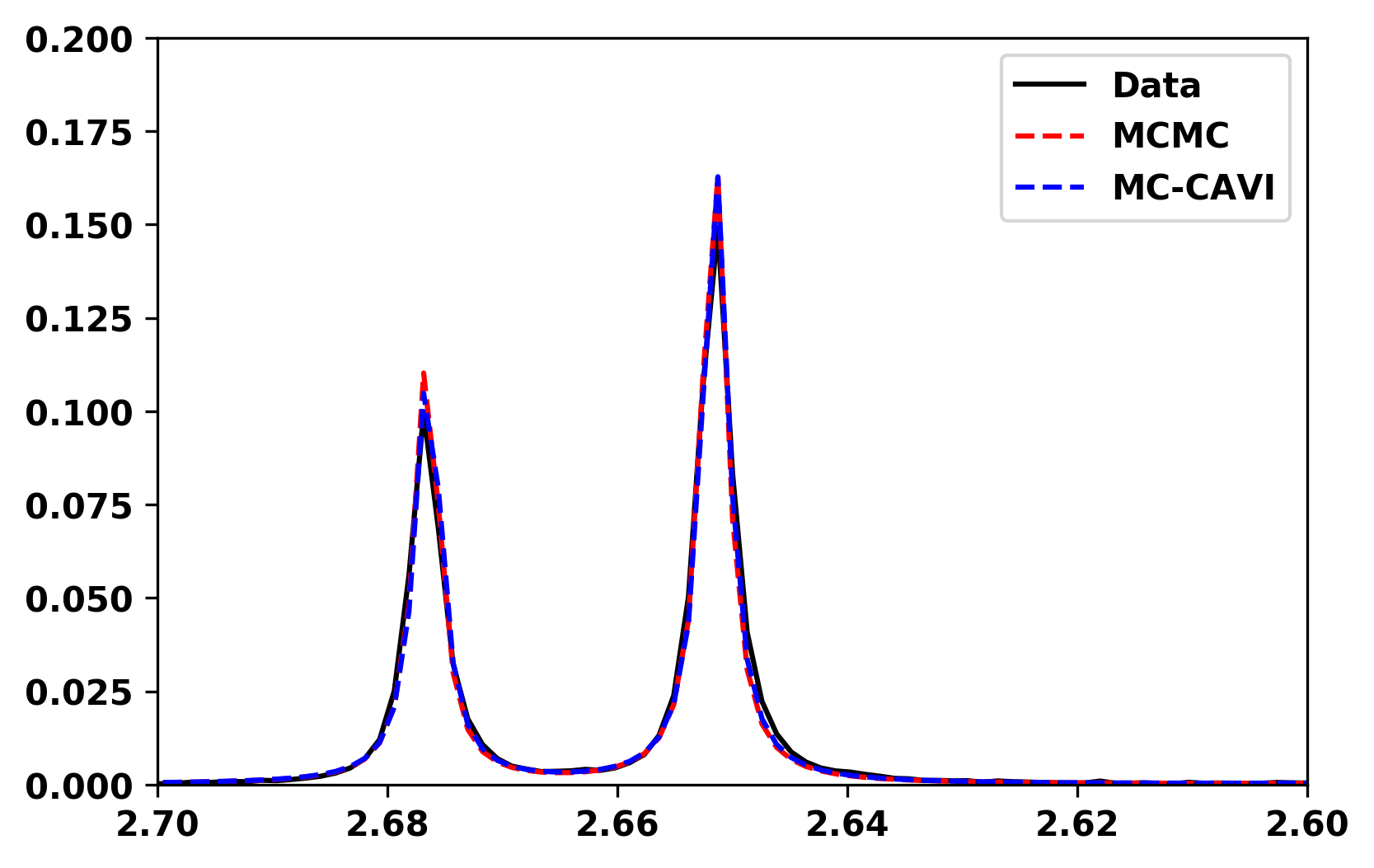}
\includegraphics[scale=0.5]{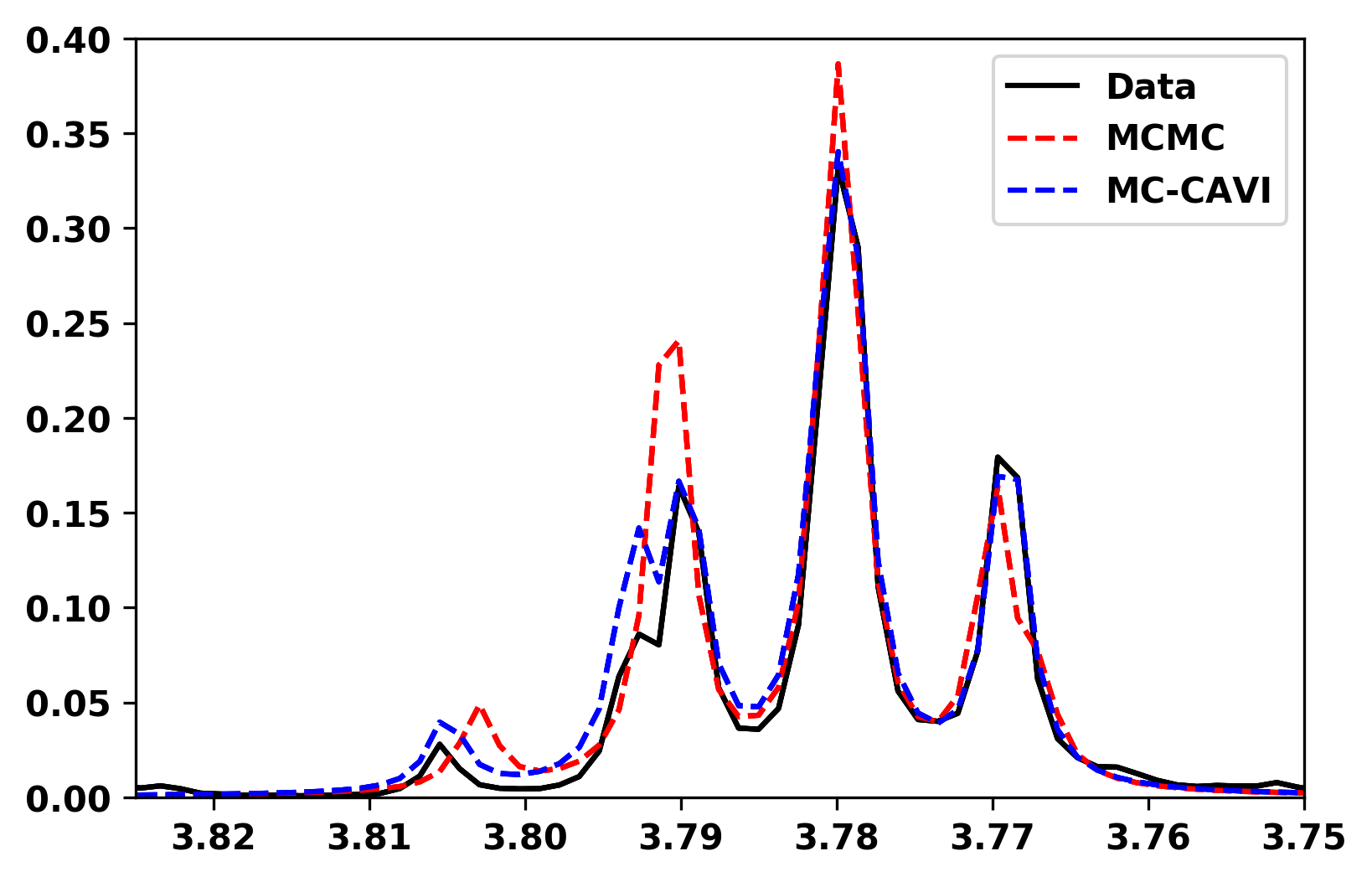}
\includegraphics[scale=0.5]{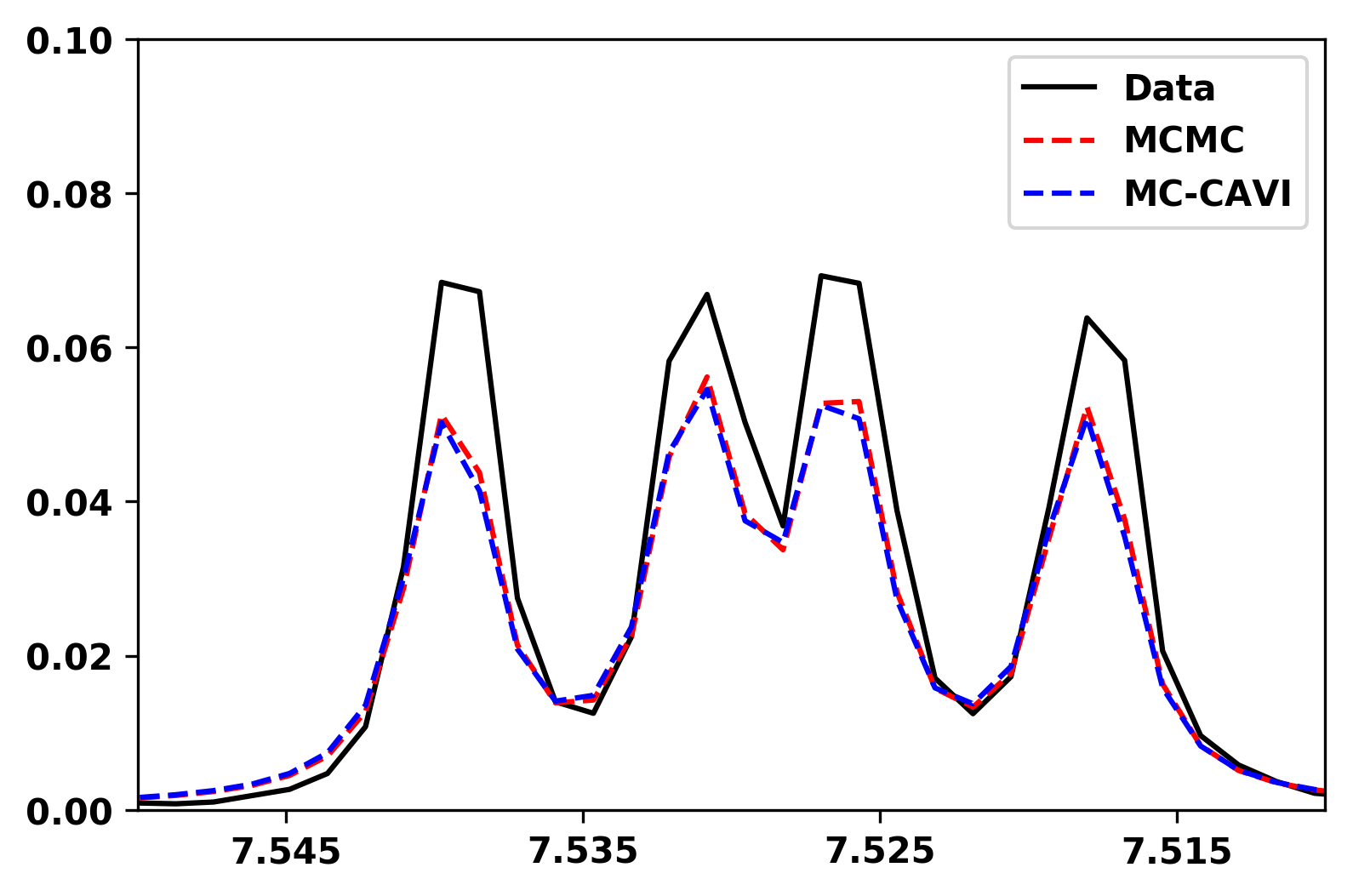}
\end{center}
\caption{Comparison of Metabolites Fit obtained with MC-CAVI and MCMC. The $x$-axis corresponds to chemical shift measure in ppm. The $y$-axis corresponds to standard density. The upper left panel shows areas around ppm value 2.14 ($\beta_4$ and $\beta_9$). The upper right panel shows areas around ppm 2.66 ($\beta_6$). The lower left panel shows areas around ppm value 3.78 ($\beta_3$ and $\beta_9$). The lower right panel shows areas around ppm 7.53 ($\beta_{10}$).}
\label{detailcomparison}
\end{figure}

In $^1$H NMR analysis, $\beta$ (the concentration of metabolites in the biofluid) and $\delta^*_{m,u}$ (the peak positions) are the most important parameters from a scientific point of view. Traceplots of four examples ($\beta_3$, $\beta_4$, $\beta_9$ and $\delta_{4,1}$) are shown in Figure \ref{paracomparison}. These four parameters are chosen due to the different performance of the two methods, which are closely examined in Figure \ref{detailcomparison}. For $\beta_3$ and $\beta_9$, 
traceplots are still far from convergence for MCMC, while they move toward the correct direction (see Figure \ref{paracomparison}) when using MC-CAVI. For $\beta_4$ and $\delta_{4,1}$, both parameters reach a stable regime very quickly in MC-CAVI, whereas the same parameters only make local moves when implementing MCMC. For the remaining parameters in the model, both algorithms present similar results.

\begin{figure}[!h]
\begin{center}
\includegraphics[scale=0.3]{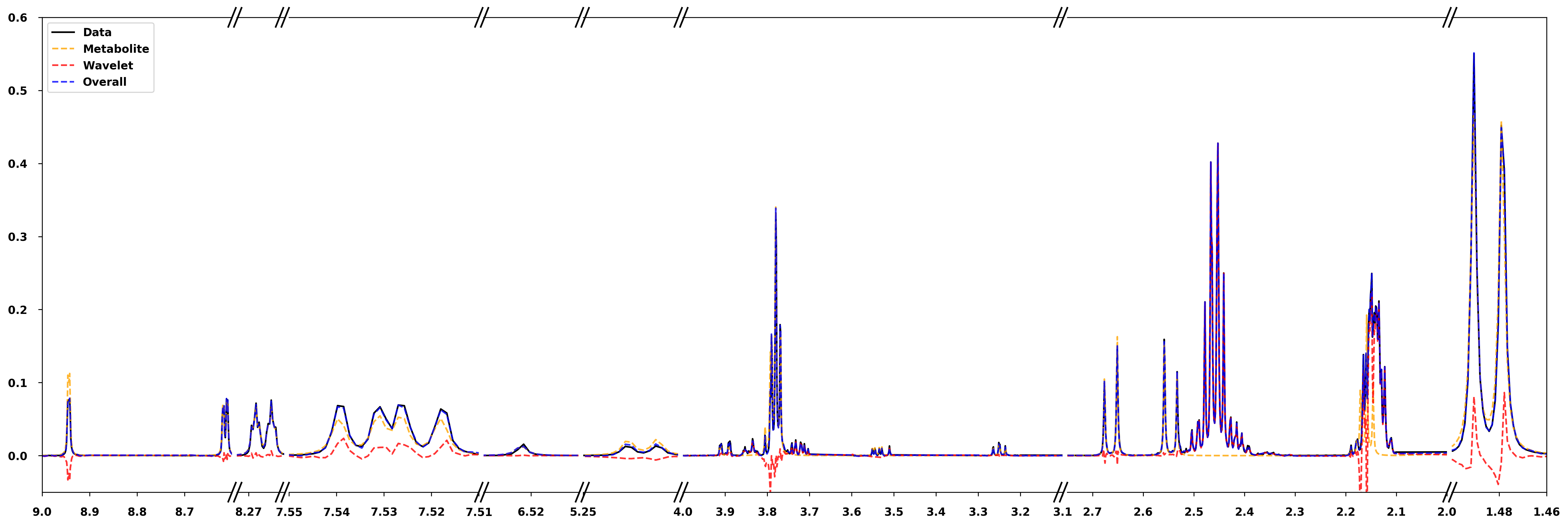}
\includegraphics[scale=0.3]{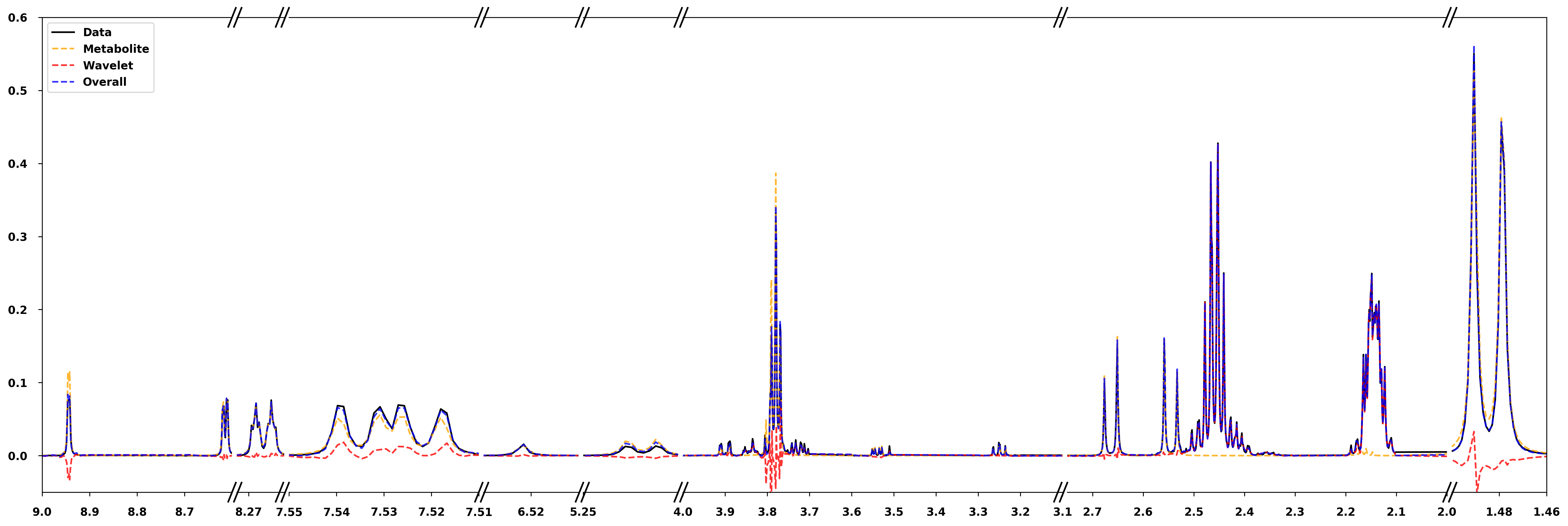}
\end{center}
\caption{Comparison of MC-CAVI and MCMC in terms of Spectral Fit. The upper panel shows the Spectral Fit from MC-CAVI algorithm. The lower panel shows the Spectral Fit from MCMC algorithm. The $x$-axis corresponds to chemical shift measure in ppm. The $y$-axis corresponds to standard density.}
\label{speccomparison}
\end{figure}

Figure \ref{speccomparison} shows the fit obtained from both the algorithms, while Table \ref{betatable} reports posterior estimates for $ \beta$.
From Figure \ref{speccomparison},  it is evident that the overall performance of MC-CAVI is similar as that of MCMC since in most areas, the metabolites fit (orange line) captures the shape of the original spectrum quite well. Table \ref{betatable} shows that, similar to standard VI behaviour, MC-CAVI underestimates the variance of the posterior density. We examine in more detail the posterior distribution of the $\beta$ coefficients for which the posterior means obtained with the two algorithms differ more than 1.0e-4. Figure \ref{detailcomparison} shows that MC-CAVI manages to capture the shapes of the peaks while MCMC does not, around ppm values of 2.14 and 3.78, which correspond to spectral regions where many peaks overlap making peak deconvolution challenging. This is probably due to the faster convergence of MC-CAVI. Figure \ref{detailcomparison} shows that for areas with no overlapping (e.g.~around ppm values of 2.66 and 7.53), MC-CAVI and MCMC produce similar results.

\begin{table}[!h]
\centering
\begin{center}
\begin{tabular}{|l|l|l|l|l|l|l|}
\hline
                         &      & $\beta_1$   & $\beta_2$   & $\boldsymbol{\beta_3}$ & $\boldsymbol{\beta_4}$ & $\beta_5$   \\ \hline
\multirow{2}{*}{MC-CAVI} & mean & 6.0e-6  & 7.8e-5  & 1.4e-3         & 4.2e-4         & 2.6e-5  \\ \cline{2-7} 
                         & sd   & 1.8e-11 & 4.0e-11 & 1.3e-11        & 1.0e-11        & 6.2e-11 \\ \hline
\multirow{2}{*}{MCMC}    & mean & 1.2e-5  & 4.0e-5  & 1.5e-3         & 2.1e-5         & 3.4e-5  \\ \cline{2-7} 
                         & sd   & 1.1e-10 & 5.0e-10 & 1.6e-9         & 6.4e-10        & 3.9e-10 \\ \hline
                         &      & $\beta_6$   & $\beta_7$   & $\beta_8$          & $\boldsymbol{\beta_9}$ & $\beta_{10}$  \\ \hline
\multirow{2}{*}{MC-CAVI} & mean & 6.1e-4  & 3.0e-5  & 1.9e-4         & 2.7e-3         & 1.0e-3  \\ \cline{2-7} 
                         & sd   & 1.5e-11 & 1.6e-11 & 3.9e-11        & 1.6e-11        & 3.6e-11 \\ \hline
\multirow{2}{*}{MCMC}    & mean & 6.0e-4  & 3.0e-5  & 1.8e-4         & 2.5e-3         & 1.0e-3  \\ \cline{2-7} 
                         & sd   & 2.3e-10 & 7.5e-11 & 3.7e-10        & 5.1e-9         & 7.9e-10 \\ \hline
\end{tabular}
\caption{Estimation of $\beta$ obtained with MC-CAVI and MCMC. (The coefficients of $\beta$ for which the posterior means obtained with the two algorithms differ by more than 1.0e-4 are shown in bold.)}
\label{betatable}
\end{center}
\end{table}

Comparing MC-CAVI and MCMC's performance in the case of the NMR model, we can draw the following conclusions:
\begin{itemize}
\item In NMR analysis, if many peaks overlap (see Figure \ref{detailcomparison}), MC-CAVI can provide better results than MCMC.
\item In high-dimensional models, where the number of parameters grows with the size of data, MC-CAVI can converge faster than MCMC.
\item Choice of $N$ is important for optimising the performance of MC-CAVI. Building on results derived for other Monte Carlo methods (e.g.~MCEM), it is reasonable to choose a relatively small number of Monte Carlo iterations at the beginning when the algorithm can be far from regions of parameter space of high  posterior probability, and gradually increase the number of Monte Carlo iterations, with the maximum number taken once the algorithm has reached a mode.
\end{itemize}

\section{Discussion}
\label{sec:discussion}
As a combination of VI and MCMC, MC-CAVI provides a powerful inferential tool particularly in high dimensional settings when full posterior inference is computationally demanding and the application of optimization and of noisy-gradient-based approaches, e.g. BBVI, is hindered by the presence of hard constraints. The MCMC step of MC-CAVI is necessary to deal with parameters for which VI approximation distributions are difficult or impossible to derive, for example due to the impossibility to derive closed-form expression for the normalising constant.  General Monte Carlo algorithms such as sequential Monte Carlo and Hamiltonian Monte Carlo can be incorporated within MC-CAVI. Compared with MCMC, the VI step of MC-CAVI speeds up convergence and provides reliable estimates in a shorter time.  Moreover, MC-CAVI scales better in high-dimensional settings. As an optimization algorithm, MC-CAVI's convergence monitoring is easier than MCMC. Moreover, MC-CAVI offers a flexible alternative to BBVI. This latter algorithm, although very general and suitable for a large range of complex models, depends crucially on the quality of the approximation to the true target provided by the variational distribution, which in high dimensional setting (in particular with hard constraints) is very difficult to assess.

\section*{Acknowledgments}
We thank two anonymous referees for their comments that greatly improved the content of the paper.
AB acknowledges funding by the Leverhulme Trust Prize.

\appendix

\section{Proof of Lemma \ref{lem:stable}}
\label{sec:lem}
\begin{proof}{Part (i)}:
For a neighborhood of $\lambda^*$, we can chose a sub-neighborhood $V$ as described in Assumption \ref{ass:M}.
For some small $\epsilon>0$, the set $V_0 = \{\lambda:\textrm{ELBO}(q(\lambda))\geq 
\textrm{ELBO}(q(\lambda^*))-\epsilon\}$ has a connected component, say $V'$, so that $\lambda^*\in V'$ and $V'\subseteq V$; we can assume that $V'$ is compact. Assumption \ref{ass:M} implies that $M(V')\subseteq V_0$; in fact, since $M(V')$ is connected and contains $\lambda^*$, we have $M(V')\subseteq V'$. This completes the proof of part (i) of 
Definition \ref{def:stable}.\\
{Part (ii)}:
Let $\lambda\in V'$. 
Consider the sequence $\{M^{k}(\lambda)\}_k$ with a convergent subsequence, $M^{a_k}(\lambda)\rightarrow \lambda_1\in V'$, for  increasing integers $\{a_k\}$. Thus, we have that the following holds, $\textrm{ELBO}(q(M^{a_{k+1}}(\lambda)))\ge  \textrm{ELBO}(q(M(M^{a_{k}}(\lambda))))\rightarrow \textrm{ELBO}(q(M(\lambda_1)))$, whereas we also have that  $\textrm{ELBO}(q(M^{a_{k+1}}(\lambda)))\rightarrow\textrm{ELBO}(q(\lambda_1))$. 
These two last limits give the implication that   $\textrm{ELBO}(q(M(\lambda_1))) = \textrm{ELBO}(q(\lambda_1))$, so that $\lambda_1=\lambda^*$.
We have shown that any convergent subsequence of $\{M^k(\lambda)\}_k$ has limit $\lambda^*$; the compactness of $V'$ gives that also $M^{k}(\lambda)\rightarrow \lambda^*$. This completes the proof of part (ii) of 
Definition \ref{def:stable}.
\end{proof}

\section{Proof of Theorem \ref{th:stable}}
\label{sec:theorem}
\begin{proof}
Let $V_1$ be as $V'$ within the proof of Lemma \ref{lem:stable}. Define $V_2 = \{\lambda\in V_1:|\lambda-\lambda^*|\ge \epsilon\}$, for an $\epsilon>0$ small enough so that $V_1\neq \emptyset$.
For $\lambda\in V_2$, we have $M(\lambda)\neq \lambda$, thus there are $\nu,\nu_1>0$ such that for all $\lambda\in V_2$ and for all $\lambda'$ with $|\lambda'-M(\lambda)|<\nu$, we obtain that  $\textrm{ELBO}(q(\lambda'))-\textrm{ELBO}(q(\lambda))>\nu_1$. 
Also, due to continuity and compactness, there is $\nu_2>0$ such that for all $\lambda\in V_1$ and 
for all $\lambda'$ such that $|\lambda'-M(\lambda)|<\nu_2$, we have $\lambda'\in V_1$.
Let $R=\sup_{\lambda,\lambda'\in V_1}\{\textrm{ELBO}(q(\lambda))-\textrm{ELBO}(q(\lambda'))\}$ and 
$k_0 = [ R/\nu_1]$ where $[\cdot]$ denotes integer part. 
Notice that given $\lambda^k_N:=\mathcal{M}_N^{k}(\lambda)$, we have that 
$\{|\mathcal{M}^{k+1}_N - M(\lambda^k_N)|<\nu_2\}\subseteq \{  \lambda^{k+1}_N\in V_1 \}$.
Consider the event $F_N=\{\lambda_{N}^{k}\in V_1\,;\,k=0,\ldots, k_0\}$. 
Under Assumption \ref{ass:technical}, we have that $\mathrm{Prob}[F_N]\ge p^{k_0}$ for $p$ arbitrarily close to 1.
Within $F_N$, we have that $|\lambda_N^{k}-\lambda^*|<\epsilon$ for some $k\le k_0$, or else $\lambda_{N}^{k}\in V_2$ for all $k\le k_0$, giving that $\textrm{ELBO}(q(\lambda^k_N))-\textrm{ELBO}(q(\lambda)) > \nu_1\cdot k_0 >R$, which is impossible.  	
\end{proof}	

\section{Gradient Expressions for BBVI}
\label{sec:gradient}
\begin{align*}
\nabla_{\alpha_{\vartheta}} \log q(\vartheta) &= (\vartheta-\alpha_{\vartheta})\cdot \exp(-\gamma_{\vartheta}),\\
%
%
\nabla_{\gamma_{\vartheta}} \log q(\vartheta) &= -\tfrac{1}{2} + \tfrac{ (\vartheta-\alpha_{\vartheta})^2}{2}\cdot \exp(-\gamma_{\vartheta}),\\
\nabla_{\alpha_{\theta}} \log q(\theta) &= \big(\gamma_{\theta} - \tfrac{\Gamma'(\exp(\alpha_{\theta}))}{\Gamma(\exp(\alpha_{\theta}))} + \log(\theta)\big)\cdot \exp(\alpha_{\theta}),\\
\nabla_{\gamma_{\theta}} \log q(\theta) &= \exp(\alpha_{\theta})-\theta\cdot\exp(\gamma_{\theta}),\\ \nabla_{\alpha_{\kappa_j}} \log q(\kappa_j,\psi_j) &= \tfrac{\kappa_j-\alpha_{\kappa_j}}{\exp(2\gamma_{\kappa_j}) }
 + \tfrac{\phi(\frac{\psi_j-\alpha_{\kappa_j}}{\exp(\gamma_{\kappa_j})})-\phi(\frac{-\psi_j-\alpha_{\kappa_j}}{\exp(\gamma_{\kappa_j})})}{\exp(\gamma_{\kappa_j})(\Phi(\frac{\psi_j-\alpha_{\kappa_j}}{\exp(\gamma_{\kappa_j})})-\Phi(\frac{-\psi_j-\alpha_{\kappa_j}}{\exp(\gamma_{\kappa_j})}))},\quad 1\leq j \leq {n}\\
\nabla_{\alpha_{\psi_j}} \log q(\kappa_j,\psi_j) &= \tfrac{\psi_j-\alpha_{\psi_j}}{\exp(2\gamma_{\psi_j}) }
 + \tfrac{ \phi(\frac{2-\alpha_{\psi_j}}{\exp(\gamma_{\psi_j})})- \phi(\frac{-\alpha_{\psi_j}}{\exp(\gamma_{\psi_j})})}{\exp(\gamma_{\psi_j})(\Phi(\frac{2-\alpha_{\psi_j}}{\exp(\gamma_{\psi_j})})-\Phi(\frac{-\alpha_{\psi_j}}{\exp(\gamma_{\psi_j})}))},\quad 1\leq j \leq {n}\\
\nabla_{\gamma_{\kappa_j}} \log q(\kappa_j,\psi_j) &= \tfrac{(\kappa_j-\alpha_{\kappa_j})^2}{\exp(2\gamma_{\kappa_j})} - 1 + \tfrac{(\psi_j-\alpha_{\kappa_j}) \phi(\frac{\psi_j-\alpha_{\kappa_j}}{\exp(\gamma_{\kappa_j})})+(\psi_j+\alpha_{\kappa_j}) \phi(\frac{-\psi_j-\alpha_{\kappa_j}}{\exp(\gamma_{\kappa_j})})}{\exp(\gamma_{\kappa_j})(\Phi(\frac{\psi_j-\alpha_{\kappa_j}}{\exp(\gamma_{\kappa_j})})-\Phi(\frac{-\psi_j-\alpha_{\kappa_j}}{\exp(\gamma_{\kappa_j})}))}, \quad 1\leq j \leq {n}\\
\nabla_{\gamma_{\psi_j}} \log q(\kappa_j,\psi_j) &= \tfrac{(\psi_j-\alpha_{\psi_j})^2}{\exp(2\gamma_{\psi_j})} - 1 + \tfrac{(2-\alpha_{\psi_j})\boldsymbol{\phi}(\frac{2-\alpha_{\psi_j}}{\exp(\gamma_{\psi_j})})+(\alpha_{\psi_j})\boldsymbol{\phi}(\frac{-\alpha_{\psi_j}}{\exp(\gamma_{\psi_j})})}{\exp(\gamma_{\psi_j})(\Phi(\frac{2-\alpha_{\psi_j}}{\exp(\gamma_{\psi_j})})-\Phi(\frac{-\alpha_{\psi_j}}{\exp(\gamma_{\psi_j})}))}, \quad 1\leq j \leq {n}.
\end{align*}
\section{MC-CAVI Implementation of BATMAN}
\label{sec:BATMAN}
In the MC-CAVI implementation of BATMAN, taking both computation efficiency and model structure into consideration, we assume that the variational distribution factorises over four partitions of the parameter vectors, $q( \beta, \delta^*,\gamma)$, $q( \vartheta, \tau)$, $q(\psi)$, $q(\theta)$. This factorization is motivated by the original Metropolis-Hastings block updates in \cite{batmanmodel}.
Let  $B$ denote the wavelet basis matrix defined by the transform $\mathcal{W}$, so $\mathcal{W}(B) = \mathbf{I}_{n_1}$. We use $v_{-i}$ to represent vector $v$ without the  $i$th component and analogous notation for matrices (resp., without the $i$th column).\\

\noindent Set $\mathbb{E}(\theta) = 2a/e$, $ {\mathbb{E}}(\vartheta^2_{j,k}) = 0$, ${\mathbb{E}}( \vartheta) = 0$, $ {\mathbb{E}}(\tau) = 0$, ${\mathbb{E}}(\mathbf{T}\beta) = \mathbf{y}$, ${\mathbb{E}}
\big((\mathbf{T} \beta)^{\top}(\mathbf{T}\beta)\big) = \mathbf{y}^{\top}\mathbf{y}$.\\

\noindent For each iteration:
\begin{enumerate}

\item Set $q(\psi_{j,k}) = \mathrm{Gamma}\big(c_j+\frac{1}{2}, \tfrac{{\mathbb{E}}(\theta) \mathbb{E}(\vartheta^2_{j,k})+d_j}{2} \big)$; calculate $\mathbb{E}(\psi_{j,k})$.
\item Set 
$q(\theta) = \mathrm{Gamma}(c,c')$,
where we have defined, 
\begin{align*}
c &= a_1+n_1+\tfrac{n}{2}, \\[0.3cm]
c'&=\tfrac{1}{2}\Big\{\sum_{j,k}\mathbb{E}(\psi_{j,k}) \mathbb{E}(\vartheta^2_{j,k})+ \mathbb{E}\big((\mathcal{W}\mathbf{y}-\mathcal{W}\mathbf{T} \beta- \vartheta)^{\top}(\mathcal{W}\mathbf{y}-\mathcal{W}\mathbf{T} \beta- \vartheta)\big)\\[-0.45cm] &\qquad\qquad\qquad\qquad\qquad\qquad\qquad +r(\mathbb{E}(\tau)-h\mathbf{1}_{n})+e\Big\};
\end{align*}
calculate $\mathbb{E}(\theta)$.
\item Use Monte Carlo to draw $N$ samples from $q( \beta,\delta^*_{m,u},\gamma)$, which is derived via (\ref{eq:recursion}) as,
\begin{align*} 
q( \beta, \delta^*,\gamma) &\propto 
\exp\Big\{-\tfrac{\mathbb{E}(\theta)}{2} \big( (\mathcal{W}\boldsymbol{T}\beta)^{\top} \mathcal{W}\boldsymbol{T}\beta - 2\mathcal{W}\boldsymbol{T}\beta(\mathcal{W}\mathbf{y}- {\mathbb{E}}( \vartheta))  \big) \Big\}\\
& \qquad \qquad \qquad \qquad\qquad\qquad \times p( \beta)p( \delta^*)p(\gamma),
\end{align*}
where $p( \beta)$, $p( \delta^*)$, $p(\gamma)$ are the prior distributions specified in Section \ref{sec:priordist}.
\begin{itemize}
\item  Use a Gibbs sampler update to draw samples from $q( \beta| \delta^*_{m,u},\gamma)$. Draw each component of $ \beta=(\beta_m)$ from a univariate normal, truncated below at zero, with precision  and mean parameters given, respectively, by
\begin{gather*}
P := s_m + {\mathbb{E}}(\theta)(\mathcal{W}\boldsymbol{T}_i)^{\top}(\mathcal{W}\boldsymbol{T}_i) ,\\[0.2cm] (\mathcal{W}\boldsymbol{T}_i)^\top(\mathcal{W}\mathbf{y}-\mathcal{W}\boldsymbol{T}_{-i}\beta_{-i}- {\mathbb{E}}(\vartheta)) {\mathbb{E}}(\theta)/P.
\end{gather*}

\item Use Metropolis--Hastings to update  $\gamma$. Propose $\log(\gamma')\sim \mathrm{N}(\log(\gamma),V_{\gamma}^2)$. Perform accept/reject. Adapt $V_{\gamma}^2$ to obtain average acceptance rate of approximately 0.45.
\item Use Metropolis--Hastings to update  $\delta^*_{m,u}$. Propose, 
$$({\delta^*_{m,u}})' \sim \mathrm{TN}(\delta^*_{m,u},V_{\delta^*_{m,u}}^2,\hat{\delta}^*_{m,u}-0.03,\hat{\delta}^*_{m,u}+0.03).$$ Perform accept/reject. Adapt $V_{\delta^*_{m,u}}^2$ to target acceptance rate 0.45.
\end{itemize}
Calculate ${\mathbb{E}}(\mathbf{T}\beta)$ and ${\mathbb{E}}\big((\mathbf{T}\beta)^{\top}(\mathbf{T}\beta)\big)$.

\item Use Monte Carlo to draw $N$ samples from $q( \vartheta, \tau)$, which is derived via (\ref{eq:recursion}) as,
\begin{align*} 
&q( \vartheta, \tau) \propto \\  &\exp\Big\{-\tfrac{\mathbb{E}(\theta)}{2} \Big(\sum_{j,k}\vartheta_{j,k}\big( (\psi_{j,k}+1)\,\vartheta_{j,k} -2\big(\mathcal{W}\mathbf{y}- \mathcal{W} \mathbb{E}
(\boldsymbol{T}\beta)\big)_{j,k} \big)  + r\sum^{n}_{i=1}(\tau_i-h)^2 \Big) \Big\}\\
&\quad\quad\quad\quad\quad \quad \quad \quad \quad \quad \quad\quad \quad \quad  \quad \quad \quad  \times \mathbb{I}\,\big\{\,\mathcal{W}^{-1} \vartheta\geq \tau,\,\, h\mathbf{1}_{n}\geq\tau\,\big\}
\end{align*}
\begin{itemize}
\item Use Gibbs sampler to draw   from $q( \vartheta|  \tau)$. Draw   $\vartheta_{j,k}$ from:
\begin{equation*} 
\mathrm{TN}\big(\tfrac{1}{1+\mathop{\mathbb{E}}(\psi_{j,k})}\big(\mathcal{W}\mathbf{y}- \mathcal{W}{\mathbb{E}}(\boldsymbol{T}\beta)\big)_{j,k},\tfrac{1}{ {\mathbb{E}}(\theta)(1+ {\mathbb{E}}(\psi_{j,k}))},L,U\big)
\end{equation*}
 where we have set, 
\begin{align*}
L &= \max_{i:B_{i\{j,k\}}>0}  \frac{\tau_i- B_{i-\{j,k\}} \vartheta_{-\{j,k\}}}{B_{i\{j,k\}}} \\
U &= \min_{i:B_{i\{j,k\}}<0} \frac{\tau_i-B_{i-\{j,k\}} \vartheta_{-\{j,k\}}}{B_{i\{j,k\}}}
\end{align*}
and $B_{i\{j,k\}}$ is the $(j,k)$th element of the $i$th column of $B$. 
\item Use Gibbs sampler to update $\tau_i$. Draw,  $$\tau_i\sim\mathrm{TN}\big(h,1/({\mathbb{E}}(\theta)r),-\infty,\min\big\{h,(\mathcal{W}^{-1} \vartheta)_i\big\}\big).$$
\end{itemize}
Calculate $ {\mathbb{E}}(\vartheta^2_{j,k})$, ${\mathbb{E}}(\vartheta)$, $ {\mathbb{E}}(\tau)$.
\end{enumerate}

\bibliography{mc-cavi}
\end{document}